\documentclass[a4paper,11pt]{article}
\usepackage{jheppub} % for details on the use of the package, please
\usepackage{physics}
\usepackage{subfigure}

\preprint{KUNS-3098}

\title{\boldmath 
Holography and Optimal Transport:\\
Emergent Wasserstein Spacetime in Harmonic Oscillator, SYK and Krylov Complexity
}

\author{Koji Hashimoto,}
\author{Norihiro Tanahashi}

\affiliation{Department of Physics, Kyoto University, Kyoto 606-8502, Japan}

\emailAdd{koji@scphys.kyoto-u.ac.jp}
\emailAdd{tanahashi@gauge.scphys.kyoto-u.ac.jp}

\abstract{
Optimal transport and Wasserstein distance are prominent tools to quantify the space of probability distributions. From a novel viewpoint of manifold hypothesis in machine learning being a possible guide for the holographic principle, we study how holographic spacetime can emerge from quantum systems in general as a Wasserstein space through optimal transport. We employ the simplest example of a single quantum harmonic oscillator and demonstrate that, among various definitions of distance, the manifold hypothesis selects the 1-Wasserstein distance of optimal transport between Husimi Q-representations of states, and it gives rise to an emergent space. 
Furthermore, the Lindblad time evolution of the harmonic oscillator coupled to a bath, of the form of a Fokker-Planck equation, provides a time trajectory in the Wasserstein space, yielding an {\it emergent Wasserstein spacetime} that shares properties with black hole spacetimes and their event horizons. 
The methodology is applied to a Lindbladian subsystem of SYK model, revealing that the Wasserstein space is consistent with the AdS${}_2$ black hole geometry of the standard holographic dictionary. 
We remark that, in our examples, the 1-Wasserstein distance is identified as a generalized Krylov complexity, and argue that optimal transport with the manifold hypothesis can yield general emergent spacetimes, positioning the holographic principle on a broader basis.
}

\begin{document}

\maketitle
\flushbottom

\section{Holography, optimal transport and manifold hypothesis}
\label{sec:intro}

The difficulty in understanding the holography \cite{hooft1993dimensional,Susskind:1994vu}, or the AdS/CFT correspondence \cite{Maldacena:1997re}, is often said to be the emergence of the gravitational spacetime out of the lower-dimensional boundary quantum field theory. The spacetime dimensions increase by one spatial dimension, acquiring the holographic dimension whose coordinate specifies the energy scale of the boundary quantum field theory. Therefore, the bulk reconstruction paradigm for obtaining the curved geometry from the information of the boundary quantum field theory has been one of the main challenges in the study of the holographic principle.

However, from the viewpoint of the fact that in the holographic principle a {\it classical} gravity emerges from the {\it quantum} nature of the boundary theory, the emergence of only the holographic dimension is intuitively rather unusual. The reason is as follows. In any quantum theory, states are described by state vectors in the Hilbert space and thus are distributions, such as wave functions. The entire space of all possible distributions is infinite-dimensional, meaning that if a quantum theory is described by a certain higher-dimensional classical theory, the dimensions should, naturally, be infinite. Thus, the holographic principle implies that the dimensions are somehow {\it reduced}, rather than emergent.

\begin{figure}[t]
\centering
    \includegraphics[height=9cm]{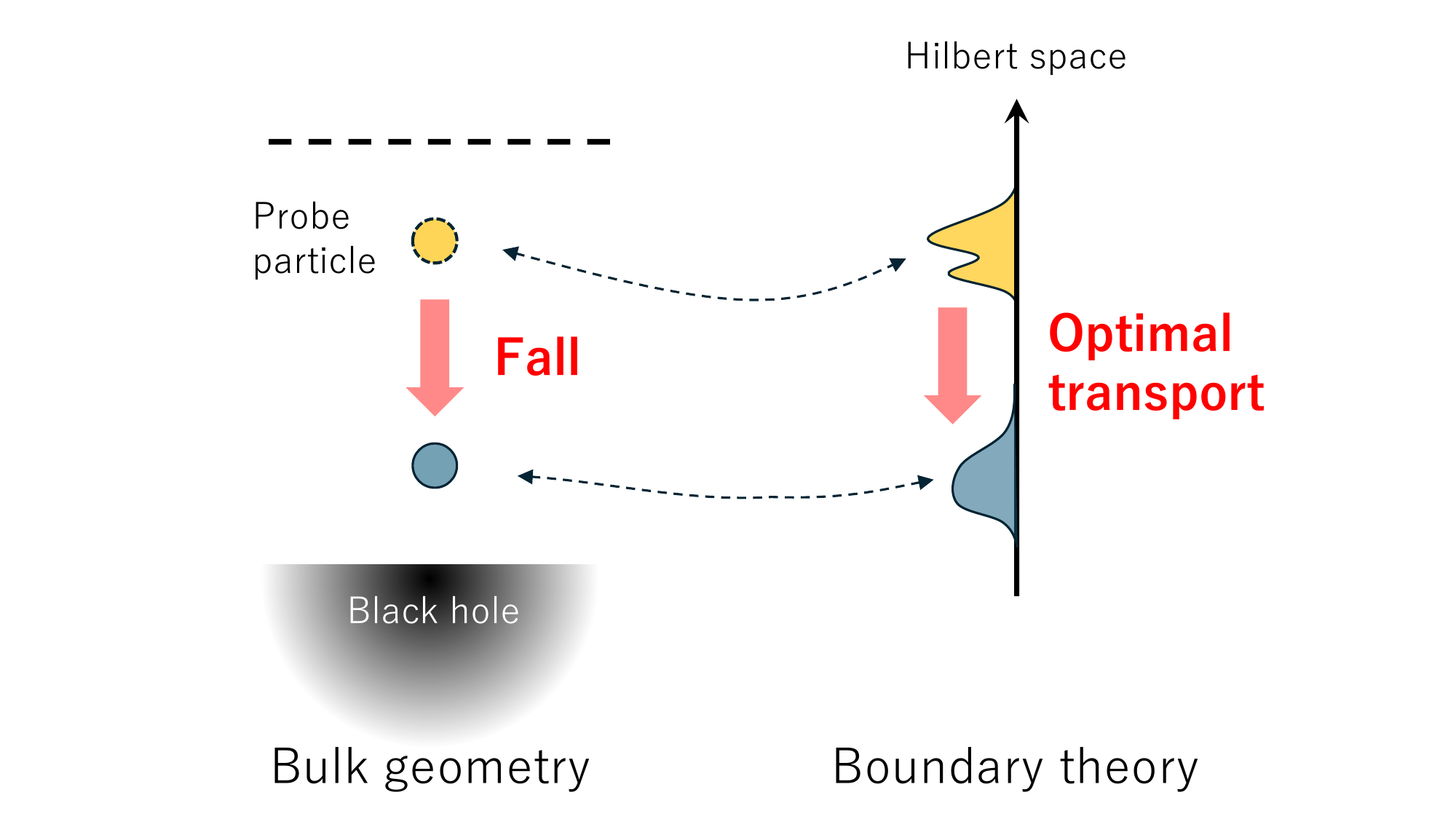}
    \caption{Schematic picture of our idea.}
    \label{fig:0}
\end{figure}

Then, how do the effective dimensions reduce? Let us consider a typical situation in holography and the AdS/CFT correspondence. To probe the gravitational curved geometry, one introduces a particle that is in motion. It may fall into a black hole. On the boundary quantum field theory side, this particle probe corresponds to a certain quantum state that is made by an operator quench. When this state is written in a Hilbert space, it is a wave function, {\it i.e.}, a distribution.
This distribution will time-evolve by the Hamiltonian of the boundary quantum field theory; thus, the motion of the probe particle and its trajectory in the curved spacetime corresponds to the change of the shape of the distributions in the Hilbert space. Now, any quantum field theory can introduce state vectors, the  dimensions of which are infinite. Furthermore, there could be a variety of distance measures introduced to the infinite-dimensional space. Therefore, from this viewpoint, we can summarize the necessary conditions for the holography to work: for a given quantum field theory, the existence of a holographic dual gravity description requires that the infinite-dimensional Hilbert space somehow reduces to an effective finite-dimensional space, with the help of a certain well-selected choice of 
\begin{description}
    \item[(i)] a distance measure, 
    \item[(ii)] a representation of the distributions,
    \item[(iii)] a peculiar set of quantum states.
\end{description}
Note that not all states in the Hilbert space allow a direct bulk interpretation in holography; thus, (iii) is needed to make a working dictionary. When all of these three are carefully prepared, a finite-dimensional bulk spacetime may be obtained.

For the mystery of the holography to be revealed, we employ the philosophy described above, and introduce two important notions from research fields related to AI and machine learning: {\it Optimal transport} and {\it Manifold hypothesis}, see Fig.~\ref{fig:0}.

\begin{description}
\item[{\bf Optimal transport}]
    provides a geometric way to compare probability measures by asking for the least-cost way to “move” one distribution into another (the Monge–Kantorovich problem). Concretely, one minimizes an average transport cost over all possible couplings $\pi (x,y)$ representing the transport between the two distributions in $x$ space and $y$ space, and the resulting optimum defines Wasserstein distances $W_p$ that metrize “shape” differences beyond pointwise comparisons. The Wasserstein distance provides a metric space, called Wasserstein space. The optimal transport thus supplies both a distance and often a transport map with a clear variational structure. A standard comprehensive reference is Villani’s monograph \cite{villani2008optimal}, while modern descriptions in connection to thermodynamics and machine learning can be found {\it e.g.} in a recent article \cite{ikeda2025speed}. 
\item {\bf Manifold hypothesis}
    posits that data living in a very high-dimensional ambient space actually concentrate near a much lower-dimensional (possibly curved) manifold, so the intrinsic degrees of freedom are far fewer than the raw coordinates suggest. This motivates nonlinear dimensionality reduction and representation learning: one seeks coordinates adapted to the manifold geometry (e.g., preserving local neighborhoods or approximate geodesic structure) rather than linear variance alone. A widely cited survey of manifold learning is Cayton’s technical report \cite{cayton2005algorithms}.
\end{description}
    
These methods can be used to select the most preferable choices of (i) distance measure and (ii) representation of the distributions. The optimal transport can provide us with $p$-Wasserstein distances, and we can seek the most preferable number $p$ $(\geq 1)$ for holography to work. The manifold hypothesis tells us that in most cases of the data sciences, the hidden effective manifolds are curved (the popular example is a ``Swiss roll'' manifold), which cautions us in selecting possible representations of the quantum states for the holography to work.  

In this paper, we conduct our research to make a possible ground for revealing the mystery of holography, with the philosophy that the holography should be a dimensional reduction mechanism from the infinite dimensional space(time)s of probability distributions of the boundary theory. The goal is to find the best choice of (i), (ii) and (iii), with which {\it the Wasserstein space is regarded as a holographic curved spacetime.}

As the simplest physical model, we study a single quantum harmonic oscillator. The harmonic oscillator can be thought of as a representative of quantum field theories, and, if our philosophy works, it could be applied to various quantum field theories. And what is more, the harmonic oscillator allows analytic treatment of its Hilbert space, which helps the tractability of the philosophy.

Then, for the choices in (i), (ii) and (iii), we employ the following strategy:
\begin{description}
\item[(i) Distance measures: $p$-Wasserstein distances.] 
As for the choices of (i) the distance measures, we employ $p$-Wasserstein distances with various $p$, rather than, for example, the Kullback-Leibler divergences, the mutual information or the Fubini-Study metric. The reason is that the $p$-Wasserstein distances are with the optimal transport: certainly, if the Wasserstein space is identified with the holographic space then a probe particle motion in the holographic dimension should follow an equation of motion which is nothing but the optimization of the particle action in the bulk. So the notion of the optimal transport is well along the idea of the holographic principle.\footnote{See the discussion in Sec.~\ref{sec:sum} for more reasons to employ the Wasserstein distances.}
\item[(iii) Sets of quantum states: Energy eigenstates.] 
As for the choices of (iii) the quantum states, we employ energy eigenstates of the harmonic oscillator. This is because in holography the holographic direction is understood as the direction of the energy; thus, to probe the holographic space, we better prepare states with different energy values. 
\item[(ii) Representations of quantum states:] {\bf Probability distributions in real space, and Husimi Q-representation in phase space.} 
With the choices given above, we apply the manifold hypothesis to choose a good representation of the quantum states for the dimensional reduction to work. As for the possible choices of (ii) the representations of quantum states, we try the ordinary probability distribution in $x$ space and Husimi Q-representation \cite{husimi1940some,takahashi1986wigner} in the phase space. Since the Husimi Q-representation is positive semidefinite and normalized, it suits optimal transport rather than using e.g.~Wigner representation, so it was used to compute the 1-Wasserstein distance for the harmonic oscillator energy eigenstates in \cite{zyczkowski1998monge}. We compare the results through the manifold hypothesis.
\end{description}

In this paper, we are going to find that the $1$-Wasserstein distance (for (i)) with the Husimi Q-representation (for (ii)) is the best choice for having the holography for the harmonic oscillator energy eigenstates.

Note that the obtained Wasserstein space does not possess time direction, while the holography should provide curved spacetimes. This motivates us to consider a motion in the Hilbert space. The energy eigenstate does not move in the Hilbert space by the Hamiltonian time evolution, so we couple the system to a bath for the states to move. The methodology is known as Lindbladian \cite{Lindblad:1975ef,gorini1976completely} with which the time evolution of states, the optimization, is described solely within the original Hilbert space. Using the Lindblad master equation applied to the harmonic oscillator with the creation/annihilation operator as a jump operator (coupling to the bath), we show that the motion in terms of the distribution is governed by the Fokker-Planck equation, whose solution provides the time evolution of the Wasserstein distance.

From the obtained Wasserstein distance, we can reconstruct the metric in the Wasserstein spacetime by regarding the Wasserstein distance as the radial direction of the holographic spacetime emergent from the harmonic oscillator.
We will find that the metric actually shares a property of a black hole horizon: the asymptotic behavior of the time evolution of the Wasserstein distance can be understood as a horizon redshift, as seen from the boundary of the spacetime. Therefore, we can claim that the emergent Wasserstein spacetime is a black hole.

All these results are for the harmonic oscillator, which is a toy model of a quantum field theory, and it is unlikely that the harmonic oscillator allows an exact gravity dual. So, we are driven to check our philosophy with popular examples in AdS/CFT correspondence: the SYK model \cite{sachdev1992gapless,kitaev2015}. 
We consider a Lindblad subsystem of a single SYK without disorder average, and applied our strategy to obtain the time evolution of the Wasserstein distance in the subsystem Hilbert space. We find that the resultant Wasserstein spacetime is identical to the AdS${}_2$ Schwarzschild geometry, where the Wasserstein distance is the radial coordinate of the geometry. So, at least in this case, we see a consistency with the popular example of the AdS/CFT correspondence.

Finally, we seek the reason why this kind of argument works. Our relevant observation is that the mathematical formula of the 1-Wasserstein distance in our case is indeed equivalent to a generalization of Krylov complexity. In fact, we can even define a ``Wasserstein operator" as an analogy of a complexity operator given in \cite{balasubramanian2022quantum}. The Krylov complexity \cite{parker2019universal,rabinovici2021operator,caputa2022geometry,balasubramanian2022quantum} was conjectured to quantify the geometric measures in the bulk \cite{Susskind:2014rva,Stanford:2014jda,Brown:2015bva,Susskind:2020gnl}. It has been studied in connection with holography in various contexts of the AdS/CFT correspondence. The observation that the 1-Wasserstein distance of ours can be equal to the generalized Krylov complexity not only supports our strategy for holography, but also suggests that there may be more hidden structure connecting the holography and the optimal transport.

The organization of this paper is as follows. First in Sec.~\ref{sec:HO1}, we apply the manifold hypothesis to the harmonic oscillator energy eigenstates and find that the $1$-Wasserstein distance with the Husimi Q-representation lets a 1-dimensional holographic dimension emerge. In Sec.~\ref{sec:HO2}, we use the Lindblad harmonic oscillator to calculate the time evolution of the 1-Wasserstein distance, and show that the bulk geometry reconstructed from the Wasserstein distance is a black hole geometry. In Sec.~\ref{sec:SYK}, we apply our strategy to the SYK model and will find that the Wasserstein distance becomes the radial coordinate of the AdS${}_2$ Schwarzschild geometry. In Sec.~\ref{sec:WK}, we argue the similarity between our 1-Wasserstein distance and the Krylov complexity, and provide more examples of Lindblad harmonic oscillators to show how far the Krylov complexity can be generalized with the idea of the optimal transport. The final section is for a summary and discussion.
Appendix provides numerical validation of the results reported in Sec.~\ref{sec:HO2}.

\section{Emergent Wasserstein space of harmonic oscillator}
\label{sec:HO1}

In this section, based on the idea of using the manifold hypothesis as a guiding principle to find a holographically emergent space, we study energy eigenstates of a quantum harmonic oscillator. 
The goal is to locate the distance measure and the representation of the quantum states that suit the holography. 
We find that the best choice is the 1-Wasserstein distance with the Husimi Q-representation.

\subsection{Choices for representations of the distributions and distances}

In general, quantum wave functions have signs and phases that cause a problem in using the standard techniques of optimal transport, as the latter is formulated for the distance measure for positive-semidefinite normalized distributions. We therefore take two kinds of distribution that are positive-semidefinite and suitable for the optimal transport: Probability distribution $\rho(x)\equiv|\braket{x}{\psi}|^2$, and Husimi Q-representation $Q(\alpha)\equiv \frac{1}{\pi} |\braket{\alpha}{\psi}|^2$ where $\ket{\alpha}$ is a coherent state.

\begin{figure}[t]
\centering
    \subfigure[Probability density.]{\includegraphics[height=4.5cm]{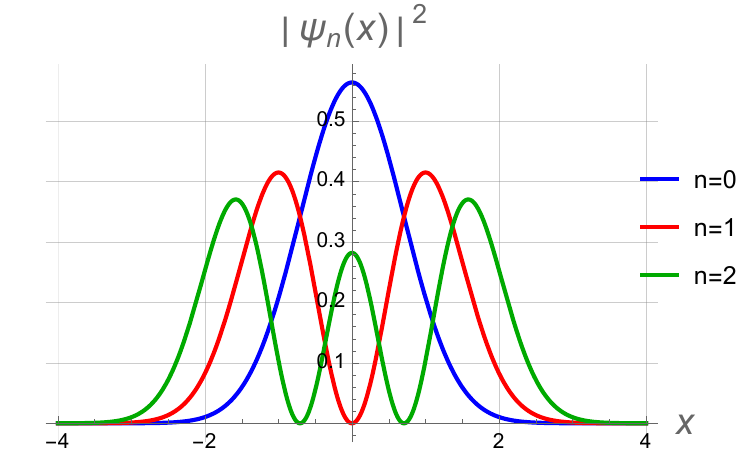}}
    \subfigure[Husimi Q-representation.]{\includegraphics[height=4.5cm]{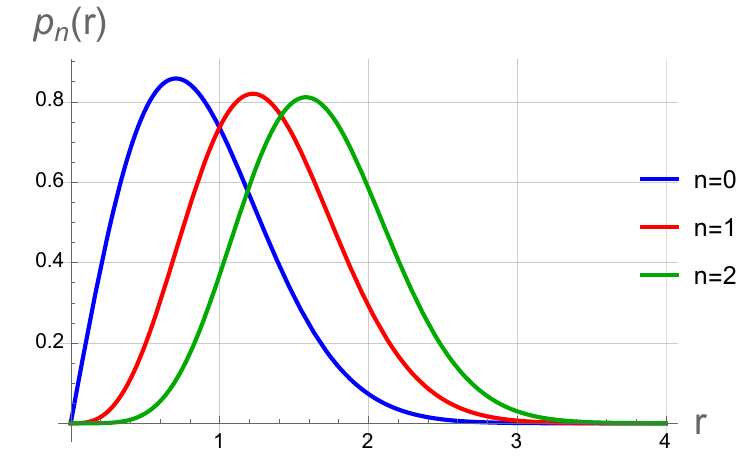}}
    \caption{The appearance of the probability density \eqref{rhon} and the Husimi Q-representation \eqref{Qn} for the harmonic oscillator energy eigenstates with $n=0,1,2$.}
    \label{fig:dist12}
\end{figure}

The single quantum harmonic oscillator, given by the Hamiltonian
\begin{align}
    \hat{H} = \frac{1}{2m}\hat{p}^2 + \frac{1}{2}m \omega^2 \hat{x}^2
\end{align}
allows the wave functions for the energy eigenstates $\ket{\psi_n}$ with energy $E_n\equiv (n+1/2)\omega$,
\begin{align}
    \braket{x}{\psi_n} = \left(\frac{m\omega}{2^{2n} (n!)^2 \pi }\right)^{1/4}
    H_n(\sqrt{m\omega}x) \exp\left[-\frac{m\omega x^2}{2}\right]
\end{align}
with the Hermite polynomial $H_n$. We take $\hbar=1$ throughout this paper.
Then the probability distribution $\rho_n(x)$ and the Husimi Q-representation $Q_n(\alpha)$ are defined as
\begin{align}
&\rho_n(x) \equiv|\braket{x}{\psi_n}|^2,
\label{rhon}
 \\
& Q_n(\alpha) \equiv \frac{1}{\pi} |\braket{\alpha}{\psi_n}|^2.
\label{Qn}
\end{align}
Here $\ket{\alpha}$ is the coherent state satisfying the eigen equation $\hat{a}\ket{\alpha}=\alpha\ket{\alpha}$ with the annihilation operator $\hat{a}\equiv (m\omega \hat{x} + i \hat{p})/\sqrt{2m\omega}$, and the complex number $\alpha$ is the eigenvalue indicating the semiclassical position in the phase space. The Husimi Q-representation of the harmonic oscillator eigenstates is
\begin{align}
    Q_n(\alpha) = \frac{1}{n! \pi} |\alpha|^{2n} e^{-|\alpha|^2}
\end{align}
which depends only on $|\alpha|$ thus is rotationally symmetric in the phase space. The distribution reduces to a 1-dimensional distribution,
\begin{align}
    p_n(r) \equiv 2 \pi r \, Q_n(r) = \frac{2}{n!}r^{2n+1}e^{-r^2}
\label{HQp}
\end{align}
defined for $r\equiv |\alpha| \, (\geq 0)$. This is a Gamma distribution.
See Fig.~\ref{fig:dist12} for the appearance of the probability distribution $\rho_n(x)$ and the Husimi representation $p_n(r)$. The former intersects with each other while the latter does not, which will be important in later analyses.

Let us turn to possible choices for the distance measure.
There is a variety of definitions for the distances between distributions, and one of the examples is the Kullback-Leibler (KL) divergence, which is commonly used in machine learning. However, the KL divergence does not provide a metric space, and is not symmetric, thus not suitable for our purpose of the emergence of space(time). We take $p$-Wasserstein distances, which are defined for any $p$   $(\geq 1)$, as they are positive semi-definite, symmetric measures and give a metric space. The Wasserstein distances are one of the popular and recent targets of research in non-equilibrium thermodynamics and machine learning such as diffusion models and Wasserstein GANs.

According to the optimal transport theory, the following $p$-Wasserstein distance solves the optimal transport problem:
\begin{align}
    W_p \equiv \left[\min_{\pi(x,y)}\int dxdy\,  \pi(x,y)\, |x-y|^p \right]^{1/p},
\end{align}
where the distance is the one between the distribution $Q_1(x)$ and $Q_2(y)$, providing the following constraint for the joint distribution $\pi(x,y)$,
\begin{align}
    Q_1(x)= \int dy \, \pi(x,y), \quad 
    Q_2(y)= \int dx \, \pi(x,y).
\end{align}
The Wasserstein distance can be defined in higher dimensions, but here in this paper we treat only 1-dimensional distributions for simplicity. 
In 1 dimension, there is a formula for the $p$-Wasserstein distances,
\begin{align}
    W_p = \left[ \int_0^1 df \left|F_{Q_1}^{-1}(f)-F_{Q_2}^{-1}(f)\right|^p\right]^{1/p}
\label{WpCDF}
\end{align}
where $F_{Q}(x)$ is the cumulative distribution function (CDF) of $Q(x)$, 
\begin{align}
    F_Q(x) \equiv \int_{-\infty}^x Q(x),
    \label{CDFdef}
\end{align}
and $F^{-1}$ is its inverse function. The integral region in \eqref{WpCDF} is from the fact that the CDF takes a value in $[0,1]$ for the normalized distribution, $F_Q(-\infty)=0$ and $F_Q(\infty)=1$. The CDF is a monotonically increasing function.

For $p=1$, i.e. 1-Wasserstein distance, the formula \eqref{WpCDF} reduces to a simpler form,
\begin{align}
    W_1 = \int_{-\infty}^\infty 
    dx \left|F_{Q_1}(x)-F_{Q_2}(x)\right|. 
\label{W1CDF}
\end{align}
That is, just the area surrounded by the two CDFs.\footnote{This is because for $p=1$ the original formula \eqref{WpCDF} is integrating the same ``region'' to measure the area in a different orientation, something like a Lebesgue integral. We would like to thank Sosuke Ito for pointing out this intuitive understanding of the Wasserstein distances.} Furthermore, for the 1-Wasserstein distance in any dimensions, the following Kantorovich–Rubinstein duality is known to hold:
\begin{align}
    W_1 = \sup_{{\rm Lip}(f)\leq 1}\left[ \int f(x) Q_1(x) dx
    -\int f(y) Q_1(y) dy  \right] 
    \label{KRd}
\end{align}
where Lipschitz seminorm Lip$(f)$ is bounded,
\begin{align}
    {\rm Lip}(f) \equiv  \frac{|f(x)-f(y)|}{|x-y|} \leq 1 .
    \label{Lip}
\end{align}
This duality, relating the 1-Wasserstein distance to an expectation value of an ``observable" $f$, plays an interesting role in the interpretation of the emergent space in the following.

In summary, for the choices of the distances, we consider $p$-Wasserstein distances \eqref{WpCDF} for various $p$, and for $p=1$ we have a simpler formula \eqref{W1CDF} for the evaluation.

\subsection{Numerical distance matrices and embedding in Euclidean space}
\label{sec:2.2}

Our principle here is, based on the manifold hypothesis, to find the best distance measure and the best representation of the quantum states. 
We are going to embed the calculated distance matrix for all pairs of energy eigenstates and will check the minimal embedding dimension $D_{\rm min}$. When this value $D_{\rm min}$ is low, the quantum states are embedded in a low-dimensional space, which is better for the holography. 

In this paper, we employ the simplest method of embedding into Euclidean spaces. 
This is because the manifold hypothesis tells that the points in Euclidean space collapse to a submanifold embedded in Euclidean space. Therefore, the first thing we need to check is whether the distance matrix made by $N$ quantum states can be embedded in $N$-dimensional Euclidean space or not (note that this check is already nontrivial). Then we will further study whether the embedding dimensions $D_{\rm min}$ further reduce irrespective of $N$, which selects our best choice of the distance measure and the representation.

Let us calculate numerically the distance matrix for pairs of the energy eigenstates $\ket{\psi_n}$ with $n=0,1,2,\cdots,N-1$, for each choice of $p$ for the $p$-Wasserstein distance and for each kind of the distributions: the probability distribution $\rho_n(x)$ and the Husimi Q-representation $p_n(x)$. For our numerical analyses, we take $N=6$ and $p=1,2,3,4$.\footnote{We examine the results for other values of $N, p$ in Appendix~\ref{appA}. We will confirm that essential parts of the claims in this section are universal for more general choices of $N,p$.}
Then we shall embed the distances in $D$-dimensional Euclidean space and see what is the minimum value of $D$, for each case. The manifold hypothesis for the holographic principle prefers $D=1$, which, as we will find, is achieved for $p=1$ with the Husimi Q-representation.

As the explicit analytic form of $\rho_n(x)$ and $p_n(x)$ is known, we shall just substitute it into the formulas of the $p$-Wasserstein distances for each choice of a pair of $n$'s, and numerically evaluate the distances. We obtain the following distance matrices. 

First, for the probability distribution $\rho_n(x)$ (in the unit $m=1$ and $\omega=1$ for our numerical purposes), the $p$-Wasserstein distances for $p=1,2,3,4$ are given by 
\begin{align}
    W_1(\rho_n,\rho_m) = \left(
\begin{array}{cccccc}
 0& 0.56419 & 0.846284 & 1.12838 & 1.33995 & 1.55152 \\
 0.56419 & 0 & 0.402302 & 0.576471 & 0.832663 & 0.999903 \\
 0.846284 & 0.402302 & 0 & 0.329381 & 0.494301 & 0.705237 \\
 1.12838 & 0.576471 & 0.329381 & 0 & 0.285632 & 0.426070 \\
 1.33995 & 0.832663 & 0.494301 & 0.285632 & 0 & 0.25567 \\
 1.55152 & 0.999903 & 0.705237 & 0.426070 & 0.25567 & 0 \\
\end{array}
\right)_{\!\!\!\!n+1,m+1}
\end{align}
\begin{align}
W_2(\rho_n,\rho_m) = 
    \left(
\begin{array}{cccccc}
 0 & 0.570438 & 0.923561 & 1.21146 & 1.46077 & 1.68381 \\
 0.570438 & 0 & 0.409037 & 0.675255 & 0.927138 & 1.14651 \\
 0.923561 & 0.409037 & 0 & 0.335537 & 0.571326 & 0.783814 \\
 1.21146 & 0.675255 & 0.335537 & 0 & 0.291247 & 0.50058 \\
 1.46077 & 0.927138 & 0.571326 & 0.291247 & 0 & 0.260847 \\
 1.68381 & 1.14651 & 0.783814 & 0.50058 & 0.260847 & 0 \\
\end{array}
\right)_{\!\!\!\!n+1,m+1}
\end{align}
\begin{align}
    W_3(\rho_n,\rho_m) = 
    \left(
\begin{array}{cccccc}
 0 & 0.575067 & 0.958331 & 1.26318 & 1.52791 & 1.76432 \\
 0.575067 & 0 & 0.414218 & 0.717353 & 0.984079 & 1.22115 \\
 0.958331 & 0.414218 & 0 & 0.340336 & 0.605219 & 0.835395 \\
 1.26318 & 0.717353 & 0.340336 & 0 & 0.295657 & 0.532523 \\
 1.52791 & 0.984079 & 0.605219 & 0.295657 & 0 & 0.26493 \\
 1.76432 & 1.22115 & 0.835395 & 0.532523 & 0.26493 & 0 \\
\end{array}
\right)_{\!\!\!\!n+1,m+1}
\end{align}
\begin{align}
    W_4(\rho_n,\rho_m) = \left(
\begin{array}{cccccc}
 0 & 0.578676 & 0.977884 & 1.29662 & 1.57177 & 1.81774 \\
 0.578676 & 0 & 0.418354 & 0.740328 & 1.01954 & 1.26732 \\
 0.977884 & 0.418354 & 0 & 0.344209 & 0.624452 & 0.869007 \\
 1.29662 & 0.740328 & 0.344209 & 0 & 0.299237 & 0.550428 \\
 1.57177 & 1.01954 & 0.624452 & 0.299237 & 0 & 0.268256 \\
 1.81774 & 1.26732 & 0.869007 & 0.550428 & 0.268256 & 0 \\
\end{array}
\right)_{\!\!\!\!n+1,m+1}
\end{align}
For the Husimi Q-representation $p_n(x)$, we find
\begin{align}
    W_1(p_n,p_m) = 
    \left(
\begin{array}{cccccc}
 0 & 0.443113 & 0.775449 & 1.05239 & 1.29472 & 1.51282 \\
 0.443113 & 0 & 0.332335 & 0.609281 & 0.851609 & 1.0697 \\
 0.775449 & 0.332335 & 0 & 0.276946 & 0.519274 & 0.737368 \\
 1.05239 & 0.609281 & 0.276946 & 0 & 0.242328 & 0.460423 \\
 1.29472 & 0.851609 & 0.519274 & 0.242328 & 0 & 0.218095 \\
 1.51282 & 1.0697 & 0.737368 & 0.460423 & 0.218095 & 0 \\
\end{array}
\right)_{\!\!\!\!n+1,m+1}
\end{align}
\begin{align}
    W_2(p_n,p_m) = 
\left(
\begin{array}{cccccc}
 0 & 0.444589 & 0.777009 & 1.05389 & 1.29613 & 1.51415 \\
 0.444589 & 0 & 0.332594 & 0.609611 & 0.851959 & 1.07006 \\
 0.777009 & 0.332594 & 0 & 0.277033 & 0.519397 & 0.737509 \\
 1.05389 & 0.609611 & 0.277033 & 0 & 0.242367 & 0.460483 \\
 1.29613 & 0.851959 & 0.519397 & 0.242367 & 0 & 0.218117 \\
 1.51415 & 1.07006 & 0.737509 & 0.460483 & 0.218117 & 0 \\
\end{array}
\right)_{\!\!\!\!n+1,m+1}
\end{align}
\begin{align}
    W_3(p_n,p_m) =
\left(
\begin{array}{cccccc}
 0 & 0.44579 & 0.778336 & 1.05519 & 1.29738 & 1.51534 \\
 0.44579 & 0 & 0.332826 & 0.609912 & 0.852283 & 1.07039 \\
 0.778336 & 0.332826 & 0 & 0.277114 & 0.519514 & 0.737642 \\
 1.05519 & 0.609912 & 0.277114 & 0 & 0.242405 & 0.460541 \\
 1.29738 & 0.852283 & 0.519514 & 0.242405 & 0 & 0.218137 \\
 1.51534 & 1.07039 & 0.737642 & 0.460541 & 0.218137 & 0 \\
\end{array}
\right)_{\!\!\!\!n+1,m+1}
\end{align}
\begin{align}
    W_4(p_n,p_m) = 
\left(
\begin{array}{cccccc}
 0 & 0.446795 & 0.779484 & 1.05634 & 1.2985 & 1.51642 \\
 0.446795 & 0 & 0.333036 & 0.610189 & 0.852583 & 1.07069 \\
 0.779484 & 0.333036 & 0 & 0.27719 & 0.519623 & 0.737769 \\
 1.05634 & 0.610189 & 0.27719 & 0 & 0.242441 & 0.460596 \\
 1.2985 & 0.852583 & 0.519623 & 0.242441 & 0 & 0.218157 \\
 1.51642 & 1.07069 & 0.737769 & 0.460596 & 0.218157 & 0 \\
\end{array}
\right)_{\!\!\!\!n+1,m+1}
\end{align}
From each of these distance matrices, 
we determine the intrinsic Euclidean embedding dimension using the classical multidimensional scaling, also known as the Torgerson--Gower method.
First we define the component-wise squared distance matrix $\Delta$ by $\Delta_{ij}\equiv (W_p)_{ij}^2$.
We introduce the centering matrix
$J \equiv I - (1/N)\mathbf{1}\mathbf{1}^\top$,
where $I$ is the $N\times N$ identity matrix and $\mathbf{1}=(1,\dots,1)^\top\in\mathbb{R}^N$.
The Gram matrix is then constructed as
\begin{align}
B \equiv -\frac{1}{2}\, J \Delta J.    
\end{align}
%If the distance matrix $W_p$ is Euclidean, the matrix $B$ coincides with the inner-product matrix of the centered configuration and is symmetric positive semidefinite.
We perform the eigenvalue decomposition
$B = V \Lambda V^\top$ where $\Lambda=\mathrm{diag}(\lambda_1,\dots,\lambda_N)$ with 
%$\lambda_1\ge\lambda_2\ge\cdots\ge\lambda_N$.
$|\lambda_1|\ge|\lambda_2|\ge\cdots\ge|\lambda_N|$.
Since $\det J=0$, one of the eigenvalues is always zero.
When there exists a negative eigenvalue in $\{\lambda_k\}_{k=1}^N$ of the Gram matrix $B$, the distance cannot be realized as a distance set of points in Euclidean space, that is, the Euclidean space embedding is impossible. On the other hand, if all of the eigenvalues are positive, it can be embedded at most in $N-1$ dimensions. 
In the latter case, the minimal embedding dimension is given by
\begin{align}
    D_{\min} \equiv \mathrm{rank}(B) = \#\{k \mid \lambda_k > 0\},
\end{align}
which is bounded as $D_{\min}\le N-1$.
The embedded coordinates in $\mathbb{R}^{D_{\min}}$ are obtained as
$X = V_+ \Lambda_+^{1/2}$, where $\Lambda_+$ contains the positive eigenvalues and $V_+$ the corresponding eigenvectors.

Given the non-negative eigenvalues $\{\lambda_k\}_{k=1}^N$,
we define the effective embedding dimension by the participation ratio
\begin{align}
    D_{\mathrm{eff}}
\equiv 
\frac{\left(\sum_{k=1}^N \lambda_k\right)^2}{\sum_{k=1}^N \lambda_k^2}.
\end{align}
This quantity provides a continuous measure of dimensionality that characterizes how many orthogonal directions contribute significantly to the embedding.
It reduces to the exact embedding dimension when the nonzero eigenvalues are degenerate, and approaches unity when a single eigenmode dominates.
In the presence of noise, small or negative eigenvalues are discarded prior to the evaluation of $D_{\mathrm{eff}}$.

\begin{table}[]
    \centering
    \begin{tabular}{c|cccccc|c|c}
 & 
 $\lambda_1$ & $\lambda_2$ & $\lambda_3$ & $\lambda_4$ & $\lambda_5$ & $\lambda_6$ &  
 $D_{\min}$ & $D_{\mathrm{eff}}$\\
 \hline
 $p=1$ & 
 1.60175 & 0.05537  & 0.00435 & 0.00030 & 0  & $-0.01571$ & -- & --\\
 $p=2$ & 
 1.92593 & 0.01801 & 0.01552 & 0.01100 & 0.00774 & 0  & 5 & 1.05481\\
 $p=3$ & 
 2.12906 & 0.02284 & 0.01136 & 0.00934 & 0.00390 & 0  & 5 & 1.04489\\
 $p=4$ & 
 2.26680 & 0.02425 & 0.00858 & 0.00549 & 0.00217 & 0  & 5 &  1.03590\\
    \end{tabular}
    \caption{The eigenvalues $\lambda_k$ and the embedding dimensions for $W_p$ of the probability distribution. For $p=1$, there exists a negative eigenvalue, thus any Euclidean embedding is not possible and $D_{\min}$ and $D_{\mathrm{eff}}$ are not defined.}
    \label{tab:ev1}
\end{table}

\begin{table}[]
    \centering
   \begin{tabular}{c|cccccc|c|c}
 & 
 $\lambda_1$ & $\lambda_2$ & $\lambda_3$ & $\lambda_4$ & $\lambda_5$ & $\lambda_6$ &  
 $D_{\min}$ & $D_{\mathrm{eff}}$\\
 \hline
 $p=1$ & 
 1.57160 & 0. & 0. & 0. & 0. & 0 & 1 & 1.00000\\
 $p=2$ & 
 1.57412 & $3.1\!\times\! 10^{-4}$ & $3.6\!\times\! 10^{-7}$  & $6.8\!\times\! 10^{-10}$ & $1.3\!\times\! 10^{-12}$  & 0  & 5 & 1.00038\\
 $p=3$ & 
 1.57640 & $5.3\!\times\! 10^{-4}$ & $-6.1\!\times\! 10^{-7}$ & $2.0\!\times\! 10^{-9}$  & $1.1\!\times\! 10^{-13}$ & 0  & --  & --\\
 $p=4$ & 
 1.57847 & $6.8\!\times\! 10^{-4}$ & $-2.0\!\times\! 10^{-6}$  & $4.0\!\times\! 10^{-9}$ & $-2.9\!\times\! 10^{-12}$  & 0   & --  & -- \\
    \end{tabular}
    \caption{The eigenvalues $\lambda_k$ and the embedding dimensions for $W_p$ of the Husimi Q-distribution. For $p=3$ and $p=4$, there exists a negative eigenvalue, thus any Euclidean embedding is not possible. Here ``0" means exact zero, while ``0." means numerically vanishing within our error analysis.}
    \label{tab:ev2}
\end{table}

See Table \ref{tab:ev1} and Table \ref{tab:ev2} for our results. The minimal embedding dimensions $D_{\min}$ are evaluated up to our numerical accuracy.
Detailed numerical analyses are reported in App.~\ref{appA}.

In Table \ref{tab:ev1}, we look at the embedding of the distance matrices for the $p$-Wasserstein distances between the probability densities. We find that $p=1$ cannot be embedded, while other $p$ need $D=N-1$ dimensions. Since the space dimensions $D_{\rm min}$ grows with the number of embedded points, it does not appear to follow the manifold hypothesis.\footnote{The effective dimensions $D_{\rm eff}$ are still close to unity, suggesting that it could be possible to embed these points in a {\it curved} 1-dimensional submanifold in the $N$-dimensional Euclidean space. We will not pursue this possibility in this paper. We shall discuss the distance on the probability distribution in Sec.~\ref{sec:sum}.}

In Table \ref{tab:ev2}, we look at the dimensions for the Husimi Q-representation. We find that for $p=1$, the Euclidean embedding looks perfect, as $D_{\min}=D_{\mathrm{eff}}=1$ up to numerical error. (This equation is analytically proven in the next subsection.) For $p=2$ the embedding is possible with $D_{\min}=N-1$ and it is not appropriate as the dimensions grow for more points. For $p>2$ the Euclidean embedding is not possible.

Thus, as a summary, from the resultant $D_{\min}$ and $D_{\mathrm{eff}}$, we find that the 1-Wasserstein distance $W_1$ for the Husimi Q-representation acquires the minimum embedding dimension, $D_{\min}=D_{\mathrm{eff}}=1$.
Thus, it is selected from our principle of the manifold hypothesis.\footnote{It should be noted that for general distributions in one dimension, the $p$-Wasserstein distance matrix can always be embedded into a Euclidean space only when $p=2$. Therefore, it is nontrivial that a certain set of distributions with $p$ other than 2 can be embedded into a Euclidean space, and at the same time, the embedding dimensions are lower than those for $p=2$. The treatment and the interpretation of the negative eigenvalues of the Gram matrix for the general cases may need more study.}
In this paper, for this reason, we shall use the 1-Wasserstein distance $W_1$ for the Husimi Q-representation.

\subsection{Emergent 1-Wasserstein space is the energy space}

\begin{figure}[t]
\centering
    \subfigure[CDF of the probability density.]{\includegraphics[height=4.5cm]{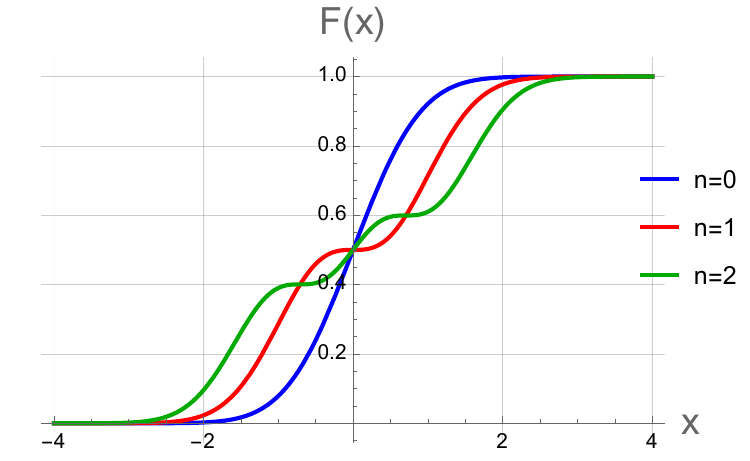}}
    \subfigure[CDF of the Husimi Q-representation.]{\includegraphics[height=4.5cm]{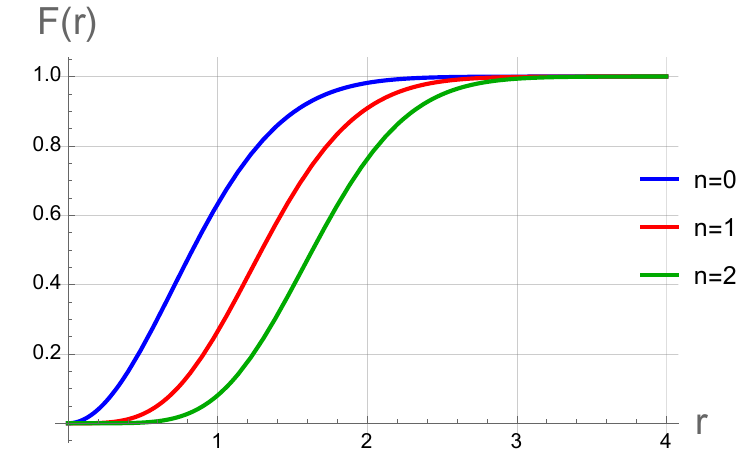}}
    \caption{The appearance of the CDFs of the probability density \eqref{rhon} and the Husimi Q-representation \eqref{Qn} for the harmonic oscillator energy eigenstates with $n=0,1,2$. The CDFs of the Husimi Q-representation are ordered, that is, they do not intersect with each other.}
    \label{fig:cdf}
\end{figure}

Let us explicitly evaluate the 1-Wasserstein distance for the Husimi Q-representation. Using the formula \eqref{W1CDF}, and noting that the CDFs of the Husimi Q-representation for the harmonic oscillator do not intersect with each other (see Fig.~\ref{fig:cdf}), we find for $n>m$
\begin{align}
    W_1(p_n, p_m) = \int_0^\infty dr \,  (F_{p_m}(r)-F_{p_n}(r))
    = Z(n)-Z(m),
    \label{znzm}
\end{align}
where through a partial integration we have defined a ``Wasserstein coordinate"
\begin{align}
    Z(n) \equiv \int_0^\infty dr \, r (p_n(r)-p_0(r)) = \langle r \rangle_{p_n}-\langle r \rangle_{p_0},
    \label{zndef}
\end{align}
where the $p_0$-related terms are the integration constant which certifies $Z(0)=0$.
Explicitly, with \eqref{HQp}, we obtain\footnote{This explicit result was first reported in \cite{zyczkowski1998monge}.}
\begin{align}
    Z(n) = \frac{\Gamma(n+3/2)}{\Gamma(n+1)} - \frac{\sqrt{\pi}}{2}.
    \label{Zn}
\end{align}
This $Z(n)$ is a monotonically increasing function, and since $W_1(n,m)$ is written as a difference between the coordinates $Z(n)$ and $Z(m)$, the Wasserstein space is exactly 1-dimensional.

The exact 1-dimensionality of this distance can be understood in two ways. 
First, through the fact that the 1-Wasserstein distance allows a simpler formula \eqref{W1CDF} and the fact that the CDFs for the Husimi Q-representation for the harmonic oscillator is completely ordered as seen in Fig.~\ref{fig:cdf}, the 1-dimensionality comes from the addition rule
 of the areas surrounded by the CDFs.
This suggests that the manifold hypothesis maximally holds when the CDFs are ordered and the 1-Wasserstein distance is employed.

The second way of understanding is through the Kantorovich-Rubinstein duality \eqref{KRd}.
The duality says that $W_1$ is provided by a difference of an expectation value of the ``operator" $f$. This operator is not an ordinary operator in quantum mechanics, in the sense that one needs to vary this operator, depending on the choice of $Q_1$ and $Q_2$, such that the difference is maximized. 
This can also be seen in the 1-Wasserstein distance \eqref{W1CDF} expressed in the following manner:
\begin{align}
    W_1(Q_1,Q_2) 
    &
    = \int_{-\infty}^\infty dx \, 
    (F_{Q_1}(x)-F_{Q_2}(x)) \, {\rm sgn} (F_{Q_1}(x)-F_{Q_2}(x))  \nonumber \\
&    = \langle f_* \rangle_{Q_1} - \langle f_* \rangle_{Q_2}
\end{align}
 where 
 \begin{align}
     f_* \equiv \int^x d\tilde{x} \, 
     {\rm sgn} (F_{Q_1}(\tilde{x})-F_{Q_2}(\tilde{x})). 
     \label{f*def}
 \end{align}
Thus, this $f_*$ is the one that satisfies the condition \eqref{Lip} and maximizes the difference of the expectation value, \eqref{KRd}. As seen from this equation \eqref{f*def}, the operator $f_*$ is $Q_1$- and $Q_2$-dependent. However, when the CDFs don't intersect with each other, the sign function reduces to a constant $1$, and the operator $f_*$ becomes state-independent: $f_* = x$. Now we see that the expression \eqref{zndef} is indeed this function.

In summary, the 1-Wasserstein distance reduces to the difference of an expectation value of a definite operator independent of the states if the CDFs are ordered. The operator in this case is $x$. In our harmonic oscillator case, this $x=r$ is $|\alpha|=(\alpha^\dagger \alpha)^{1/2}$. Since $\alpha$ is the eigenvalue of the annihilation operator, semi-classically it is 
\begin{align}
    |\alpha| \sim 
    \left|\frac{m \omega x + i p}{\sqrt{2mw}}\right|
    = \left[\frac{p^2}{2m\omega}+\frac12 m \omega x^2\right]^{1/2}
    = \left[\frac{\rm Energy}{\omega}\right]^{1/2}.
\end{align}
Therefore, semiclassically ($n \gg 1$), we find that the emergent coordinate in the 1-Wasserstein space is\footnote{Note that this expression is a weighted integration with the Husimi Q-representation. Thus it is not equal to $\langle n|H^{1/2}|n\rangle/\sqrt{\omega}$. The coherent state space spanned by $\{|\alpha\rangle\}$ is overcomplete, while one still keeps a completeness-like expansion $1 = \int d^2\alpha \, |\alpha\rangle\langle \alpha | / \pi$.\label{footH}}
\begin{align}
    Z(n) \sim \left\langle \left[\frac{\rm Energy}{\omega}\right]^{1/2}\right\rangle_{p_n}.
\end{align}
This means that the emergent 1-Wasserstein space is the energy space.

Finally, let us construct a metric of the emergent 1-Wasserstein space. Extending the energy label space $\{n\}$ (which is discrete) to a continuous 1-dimensional space with a coordinate $\zeta$, such that $\zeta=n$ corresponds to the energy eigenstate $|n\rangle$, we find that the consistent metric on the space should satisfy
\begin{align}
    Z(n)-Z(m) = \int_m^n \sqrt{g_{\zeta\zeta}(\zeta)}
\end{align}
for $n\geq m$.
Differentiating this expression leads to
\begin{align}
    g_{\zeta\zeta}(\zeta)
    = \left(\frac{d}{d\zeta}
    \frac{\Gamma(\zeta+ 3/2)}{\Gamma(\zeta+1)}
    \right)^2. 
    \label{metgzz}
\end{align}
The semiclassical approximation ($n\gg 1$) of this metric is
\begin{align}
    g_{\zeta\zeta} \sim \frac{1}{4\zeta},
    \label{gzz}
\end{align}
which comes from 
\begin{align}
    Z(n)\sim \sqrt{n} \qquad (n\gg 1)
    \label{znsqrtn}
\end{align}

Interestingly, the metric expression \eqref{gzz} diverges at $\zeta=0$, which reminds us of a black hole horizon. However, to identify it with a black hole horizon, we need to include time direction in the emergent space to make it a spacetime. In the next section, we introduce a time dependence as a motion in the Hilbert space and will find the emergence of a black hole spacetime.

%%%%%%%%%%%%%%%%%%%%%%%%%%%%%%%%%%%%%%%%%%%%%%%%%%%%%%%%%%%%%%%%%%%%%%%%%%%

\section{Emergent Wasserstein spacetime of Lindblad harmonic oscillator}
\label{sec:HO2}

In this section, we study the time evolution of quantum states of the harmonic oscillator measured by the Wasserstein distance. Following our analysis in the previous section, we shall use the 1-Wasserstein distance and the Husimi Q-representation. To make the system time-evolve, we couple the harmonic oscillator to a bath, and describe the system with the Lindblad master equation. From the growth of the Wasserstein distance of the Lindblad quantum state, we construct a metric of an emergent spacetime. The resultant metric is found to be that of a black hole with an event horizon. 

\subsection{Lindblad harmonic oscillator gives time evolution of 1-Wasserstein}

The time evolution among energy eigenstates is available once an interaction with another system is introduced. The Lindblad formulation is a popular method for describing the subsystem without any detailed knowledge of this bath system. One only specifies the jump operator, which represents the bath coupling, then the Lindblad master equation dictates the time-evolution of the subsystem due to the coupling to the bath.

We choose the annihilation operator $\hat{a}$ as the jump operator, for two reasons; first, it shows the energy flow to the bath, as the subsystem energy decreases due to the action of the jump operator, and second, the subsystem time-evolution is solvable.
The Lindblad master equation for the density matrix $\hat{\rho}(t)$ is
\begin{align}
    \frac{d}{dt}\hat{\rho} = - i \omega [\hat{n},\hat{\rho}]
    + \gamma \left(
    \hat{a}\hat{\rho}\hat{a}^\dagger - \frac12 \{ \hat{n}, \hat{\rho}\}
    \right).
    \label{Lind}
\end{align}
Here $\gamma$ is a positive constant which characterizes the coupling to the bath system. The first term on the right-hand side is the ordinary Hamiltonian time evolution term, and the second term arises due to the coupling to the bath system, using the jump operator $\hat{a}$. The last term is to make sure that the probability of the subsystem is conserved: ${\rm Tr}\hat{\rho}(t)=1$.

Due to the special form of the jump operator, we find that the time evolution of states is closed among decohered states, meaning that the following form of the density matrix
\begin{align}
    \hat{\rho}(t) = \sum_{k} C_k(t) |k\rangle \langle k |
    \label{dent}
\end{align}
becomes a solution of the Lindblad master equation if
\begin{align}
    \dot{C_k} = \gamma \left((k+1)C_{k+1}- k C_k\right).
    \label{Ck}
\end{align}

\begin{figure}[t]
\centering
    \subfigure[Husimi Q-representation.]{\includegraphics[height=4.5cm]{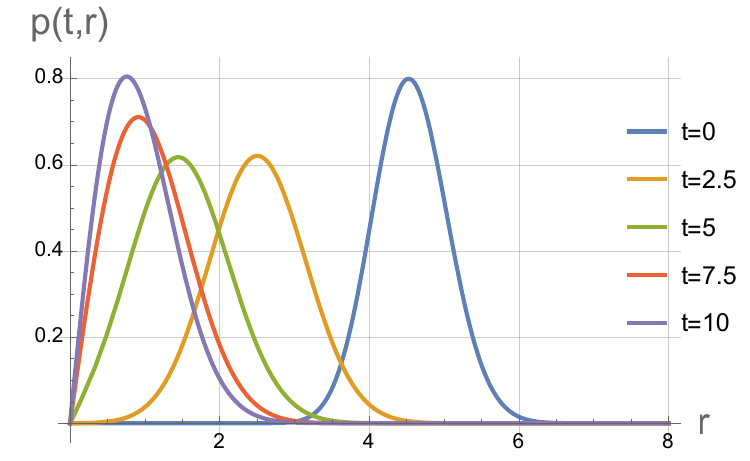}}
    \subfigure[CDF.]{\includegraphics[height=4.5cm]{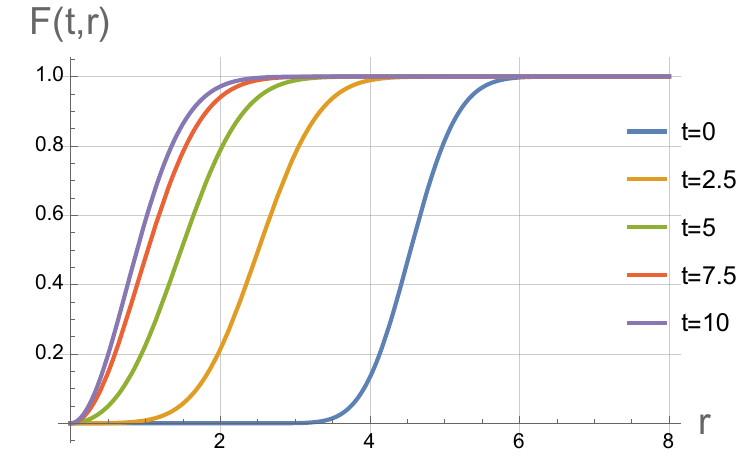}}
    \caption{Time evolution of the Lindblad harmonic oscillator, starting with $\hat{\rho}=|m\rangle\langle m|$ with $m=20$. We took the unit $\gamma=1/2$. The left panel is the time evolution of the Husimi Q-representation, and the right panel is that of the CDF.}
    \label{fig:time}
\end{figure}

Let us consider the time evolution of a state which is initially at $|m\rangle$ with $m\in \mathbf{Z}_+$. The explicit solution of \eqref{Ck} is 
\begin{align}
&    C_k(t) = 
    {}_m {\rm C}_{m-k} \, e^{-m \gamma t} \, \left(e^{\gamma t}-1\right)^{m-k} & (k\leq m) 
    \label{Cksol}\\
& C_{k}(t)=0 & (k>m)
\end{align}
Using this, we obtain the time evolution of the Husimi Q-representation as
\begin{align}
    p(t,r) = \sum_{k=0}^m
    C_k(t) \frac{2}{k!} r^{2k+1} e^{-r^2} .
    \label{ptr}
\end{align}
A numerical plot of the time evolution \eqref{ptr} is shown in Fig.~\ref{fig:time}. It shows that, at time $t=0$ the Husimi Q-representation of $|m\rangle$ is almost of the form of a Gaussian, and starts to ``fall" with a certain velocity toward $r=0$. Then it finally approaches the distribution of the ground state $|0\rangle$.
This is like a Gaussian wave packet moving toward $r=0$ and asymptotically stops. As the CDFs of this time series are completely ordered, we can compute the 1-Wasserstein distance $W_1(t)$ between $p(t,r)$ and $p(t=0,r)$ straightforwardly as
\begin{align}
    W_1(t) &= \int_0^\infty dr \left(
    F(t,r)- F(t=0,r)
    \right) \nonumber \\
    &= \int_0^\infty dr \left(
    \sum_{k=0}^m C_k(t) F_{p_k}(r) - F_{p_m}(r)
    \right) \nonumber \\
    &= \int_0^\infty dr \left(
    \sum_{k=0}^m C_k(t) F_{p_k}(r) - \left(\sum_{k=0}^m C_k(t) \right)F_{p_m}(r)
    \right) \nonumber \\
    &= \sum_{k=0}^m C_k(t) \int_0^\infty dr \left(
    F_{p_k}(r) - F_{p_m}(r)
    \right) \nonumber \\
    &= \sum_{k=0}^m C_k(t) W_1(p_k, p_m).
    \label{W1W1}
\end{align}
Using the explicit expression for $W_1$ given in \eqref{znzm} and $\sum_k C_k(t)=1$, 
we find 
\begin{align}
    W_1(t) = Z(m)-\sum_{k=0}^m C_k(t) Z(k).
    \label{W1tzm}
\end{align}

\begin{figure}[t]
\centering
    \includegraphics[height=6.5cm]{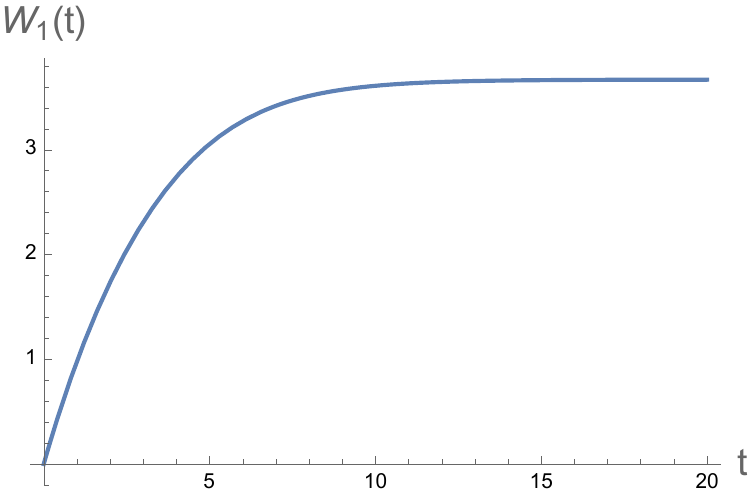}
    \caption{Time evolution of the 1-Wasserstein distance between the Husimi Q-representation of $\hat{\rho}(t)$ and $\hat{\rho}(t=0) = |m\rangle\langle m|$ with the initial condition $m=20$.}
    \label{fig:W1t}
\end{figure}

A numerical plot of this $W_1(t)$ for $m=20$ as an example is shown in Fig.~\ref{fig:W1t}.
As seen in the numerical plot, the 1-Wasserstein distance increases monotonically over time and approaches the asymptotic value $Z(m)$. Typically, at early times the 1-Wasserstein distance grows linearly in time, and at late times it approaches the final value exponentially. Let us analyse this behavior analytically. 

First, at late times, as the time-evolution solution \eqref{Cksol} shows, the dominant terms are just $C_0$ and $C_1$, because the other components are suppressed as ${\cal O}(e^{-2 \gamma t})$. From these leading terms we find
\begin{align}
    p(t,r) = 2r e^{-r^2}\left[
    (1-m e^{-\gamma t}) + mr^2 e^{-\gamma t}
    \right] + {\cal O}(e^{-2\gamma t}),
\end{align}
then the 1-Wasserstein distance is evaluated as
\begin{align}
    W_1(t) = Z(m) - \frac{\sqrt{\pi}}{4}m e^{-\gamma t}+ {\cal O}(e^{-2\gamma t}).
    \label{W1late}
\end{align}
Thus, the distance approaches the asymptotic value $Z(m)$ from below exponentially.

Second, at early times, we analyze the $p$-Wasserstein distance at $m\gg 1$, since for this choice of $m$ the Husimi Q-representation becomes almost Gaussian; 
\begin{align}
    p_m(r) \sim \sqrt{\frac{2}{\pi}} \exp[-2 \left(r-\sqrt{m+1/2}\right)^2].
    \label{Gauss}
\end{align}
In fact, a straightforward calculation shows that for small values of $t$ we find that the Gaussian distribution is moving with a constant velocity,
\begin{align}
    p(t,r) = \sqrt{\frac{2}{\pi}} \exp[-2 \left(r-\sqrt{m+1/2}+ v t\right)^2] + {\cal O}(t^3),
\end{align}
with the velocity value
\begin{align}
    v \equiv \frac12 \sqrt{m}\gamma .
\end{align}
This motion is consistent with the numerical analysis of the time evolution of the Husimi Q-representation, see Fig.~\ref{fig:time}.
The $p$-Wasserstein distance between Gaussian distributions is well-known, and we find
the early-time behavior for $t\ll 1/\gamma$ and for a large $m$,
\begin{align}
    W_p(t) \sim \frac12 \sqrt{m}\gamma t.
    \label{earlyWp}
\end{align}
So this is a linear growth.

\begin{figure}[t]
\centering
    \includegraphics[height=6.5cm]{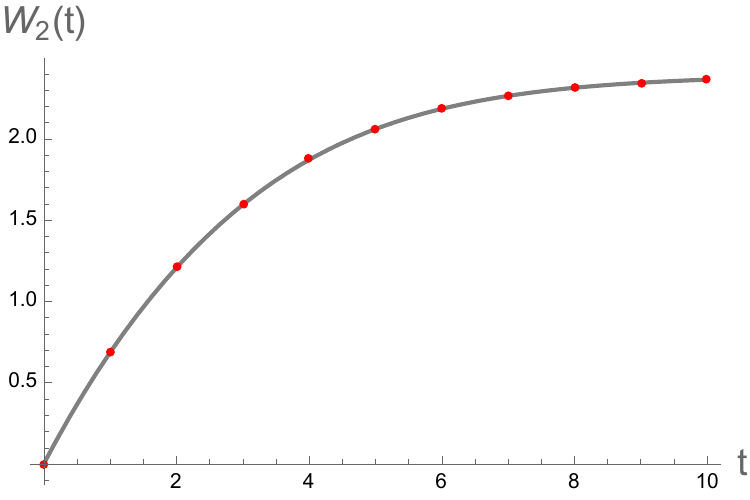}
    \caption{Red dots: time evolution of the 2-Wasserstein distance between the Husimi Q-representation of $\hat{\rho}(t)$ and $\hat{\rho}(t=0) = |m\rangle\langle m|$ with the initial condition $m=10$. Gray solid line: that of the 1-Wasserstein distance, \eqref{W1tzm}.}
    \label{fig:W2t}
\end{figure}

We can further evaluate the whole time evolution of the $p$-Wasserstein distance numerically. See Fig.~\ref{fig:W2t} for the time evolution of the 2-Wasserstein distance. We find that the results for $p=1$ and $p=2$ overlap, and numerically the difference is only $\sim 1 \%$. This is also the case for $p=3, 4$, thus we conclude that the time evolution of the $p$-Wasserstein distance is
characterized by the one with $p=1$.

\subsection{Emergent spacetime for fixed initial position}

In the previous subsection we have evaluated the time evolution of the 
1-Wasserstein distance, and described a numerical plot of it in Fig.~\ref{fig:W1t}. 
In this subsection, we aim, for a given time evolution of the Wasserstein distance, to obtain a spacetime that reproduces the time evolution.

More precisely speaking, we consider the following situation.
First, we regard the emergent curved spacetime as the Wasserstein space, in the sense that
the radial coordinate $z$ of this curved spacetime is given by the Wasserstein distance $W_p$:
\begin{align}
    z = W_p.
\end{align}
The surface $W_p=0$ denotes the boundary of the spacetime.
Suppose that one shoots a light particle from the boundary of the curved spacetime, at time $t=0$, and that it approaches a null ray or a null shock wave. Then at time $t$, the boundary quantum state should correspond to the bulk particle state in which the bulk particle is at $z=W_p$.\footnote{
From the observational point of view, there is another possibility to build the bulk spacetime. 
When the null shock wave comes to the radial point $W_p$, one tries to observe it from the boundary: the particle emits light toward the boundary, and when the light reaches the boundary at time $t_o$, we say that the particle is observed at the position $W_p$. This specifies the observation function $W_p(t_o)$. 
In the conformal coordinate, we see $t_o$ is twice as large as the time $t$ of the bulk point at which the light is emitted. Only with this factor 2 difference, the following calculations go through.}

We consider the following metric of the emergent spacetime,
\begin{align}
    ds^2 = \sigma(z) \left(-f(z) dt^2 + \frac{1}{f(z)} dz^2\right),
\end{align}
where the radial coordinate $z$ will be identified with the Wasserstein distance $W_p$,
and the ``conformal factor" $\sigma(z)$ will not be determined due to our setting; a null trajectory is insensitive to the conformal factor $\sigma(z)$. Note that the metric form includes AdS black holes, with $f(z)$ being the blackening factor and the choice $\sigma(z)=1/z^2$ bringing us to an AdS geometry.

Since the trajectory in the bulk spacetime is a null ray, it is better to make a coordinate change to $y=y(z)$, which gives a conformally flat metric
\begin{align}
    ds^2 = \sigma(z(y)) f(z(y)) (-dt^2 + dy^2),
\end{align}
through 
\begin{align}
    f(z) = \frac{dz}{dy}.
\end{align}
The surface $y=0$ is the boundary of the spacetime.
The null ray of the particle, shot at $t=0$, propagates on 
%the null cone $s=t$.
the null trajectory $y=t$.
This means the relation between the observation time and the radial position:
\begin{align}
    z(y) = W_p(t=y).
    \label{ident}
\end{align}
This determines the metric, since we have
\begin{align}
    f(z(y)) =
    %\frac{d}{ds}W_p(2y).
    \frac{d}{dy}W_p(t=y).
    \label{givef}
\end{align}

Let us use the early time and late time behavior of $W_1(t)$ to look at the characteristics of the emergent metric. First, at early time, the linear growth \eqref{earlyWp} with \eqref{givef} shows
\begin{align}
    f(z) \sim \frac12 \sqrt{m}\gamma \qquad (z\sim 0)
\end{align}
So the metric near the boundary is \text{blue}{conformally flat,}
\begin{align}
    ds^2 = \sigma(z) \left[
    -\frac{\sqrt{m}\gamma}{2} dt^2 + \frac{2}{\sqrt{m}\gamma}dz^2
    \right].
    \label{nearbm}
\end{align}
Second, at late time, the exponential approach formula \eqref{W1late} with \eqref{givef} shows
\begin{align}
        f(z) \sim \frac{\sqrt{\pi}}{4}m \gamma e^{-\gamma y}
    =\gamma \Bigl(W_1(t=\infty)-W_1(t)\Bigr)
    =\gamma(z_*-z) \qquad  (z\sim z_*) 
\end{align}
where $z_* \equiv W_1(t=\infty) = Z(m)$.
Therefore, near $z=z_*$ the metric is
\begin{align}
    ds^2 = \sigma(z) \left[
    -\gamma (z_*-z)dt^2 + \frac{1}{\gamma(z_*-z)}dz^2
    \right].
    \label{W1bh}
\end{align}
This is a black hole horizon. The horizon is located at $z=z_*=W_1(t=\infty)=Z(m)$.

Why have we obtained the black hole geometry? The following could be a phenomenological answer: When one throws a particle into a black hole while observing it from a distance, one sees that the particle slows down and almost stops near the horizon, due to the redshift. The evolution form of the Wasserstein distance shown in Fig.~\ref{fig:W1t} is exactly like that. 

Then, intrinsically, why has the black hole geometry appeared? Let us remind ourselves of the fact that we started with the Lindbladian system. We coupled the harmonic oscillator to a bath, and then kept only the bath coupling while integrating out the bath degrees of freedom, assuming the Markovian time evolution. Thus, the total system is the single harmonic oscillator and the bath. The time evolution shows that even starting with the highly excited state in the subsystem, the energy flows out to the bath and the subsystem relaxes. This is similar to what we expect for a holographic description of a black hole: the black hole degrees of freedom (which are the bath) are hidden behind the horizon, and the particle is coupled to the bath through the gravitational interaction. The potential energy of the particle decreases and finally the particle joins with the black hole. Therefore, naively these two pictures of the Lindbladian harmonic oscillator and a particle in a black-hole-like geometry share the same physical properties; thus, we conclude that the emergence of the black hole geometry is natural once a proper identification of the distribution species and the distance measure is employed.

A caveat
should be mentioned here. Of course, we are not claiming here that this Lindblad harmonic oscillator is completely dual to a gravitational system with the geometry given by the metric \eqref{W1bh}. The harmonic oscillator is unlikely to possess its gravity dual. Other quantum states of this quantum system may not be interpreted nicely with the geometric picture. Even some inconsistency, such as the fact that the Lindblad master equation is with the zero temperature while the black hole may generally have nonzero temperature,\footnote{
  Note that the coupling constant $\gamma$ in the Lindblad master equation plays the role of an effective surface gravity ($\kappa = \gamma/2$ if $\sigma(z)\sim\text{const.}$ near horizon) of the emergent black hole geometry, as seen from the near-horizon metric~\eqref{W1bh}.
  }
suggests that the ``duality" cannot be taken seriously;
  rather, in this section we just show that a black-hole-like geometry is deduced through our proposed procedures of the optimal transport and the manifold hypothesis applied to a simple quantum harmonic oscillator.\footnote{See Sec.~\ref{sec:unique} for more calculations of the Wasserstein space with a general Lindblad master equation with nonzero temperature.}

\subsection{Emergent spacetime from set of trajectories}

In the previous subsection, we constructed an emergent spacetime by a single time trajectory of the 1-Wasserstein distance of the harmonic oscillator state initially at $|m\rangle$. In this subsection, we look for the possibility of having a consistent emergent geometry for any trajectory starting with any $m$. 

\begin{figure}[t]
\centering
    \subfigure[1-Wasserstein from the shared start.]{\includegraphics[height=4.5cm]{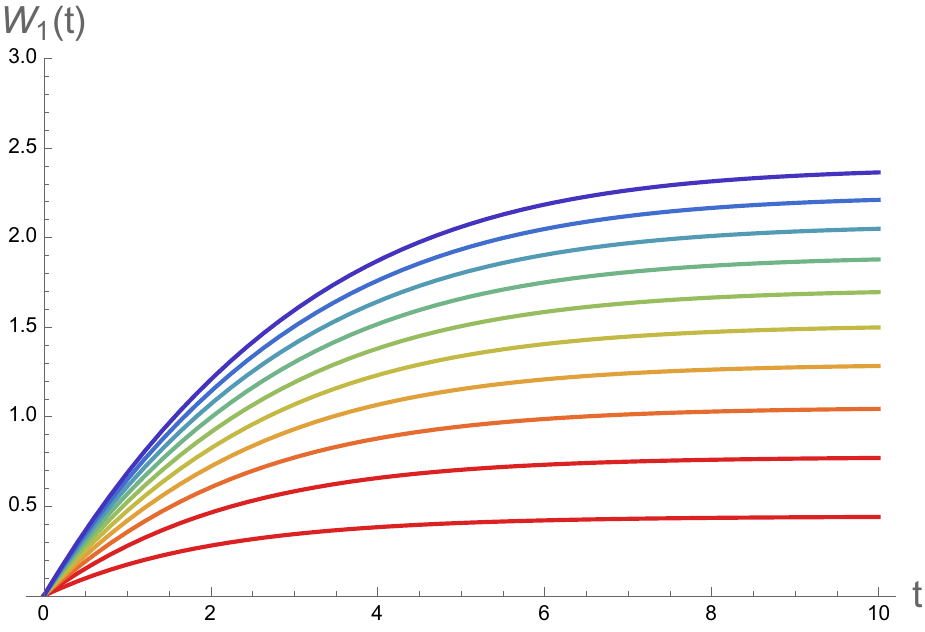}\label{figure6a}}
    \subfigure[1-Wasserstein whose origin is shifted.]{\includegraphics[height=4.5cm]{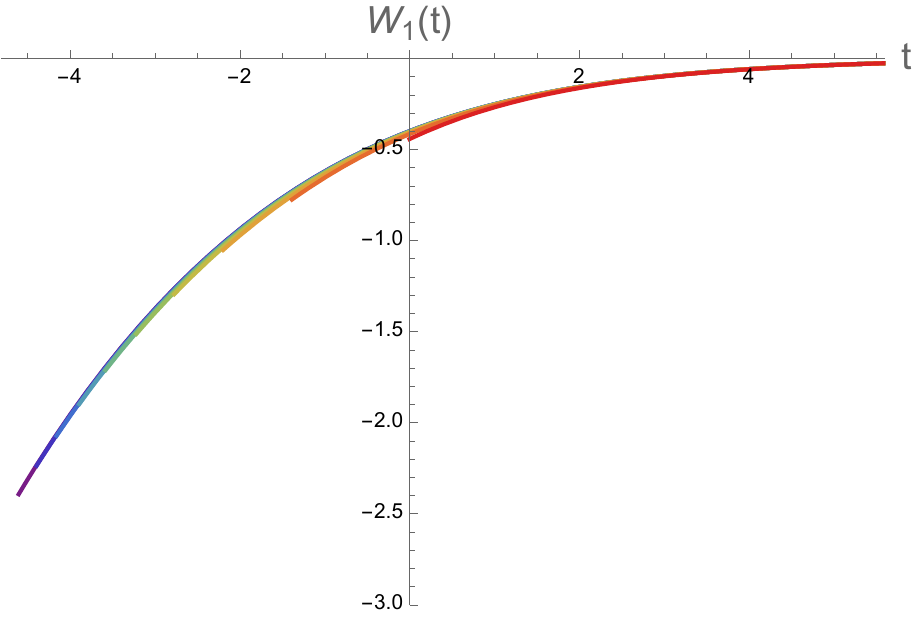}\label{figure6b}}
    \caption{Time evolution of the 1-Wasserstein distance of the Lindblad harmonic oscillator, starting with $m=1,2,\cdots,10$. In the left panel, we plot them with different colors for $m=1,2,\cdots,10$ (from bottom to top). In the right panel, we plot the same with a shifted origin. All of them roughly overlap with each other.}
    \label{fig:time2}
\end{figure}

There are infinitely many choices for the starting point $|m\rangle$. In Fig.~\ref{fig:time2}, we plot the time evolution of the 1-Wasserstein distance \eqref{W1tzm} for $m=1,2,\cdots,10$. Obviously, depending on the initial state $|m\rangle$, the curves are different. The question is whether these different trajectories are regarded as ones in a single shared emergent spacetime with different starting points in the bulk.

To see this, we look at the late time behavior of the 1-Wasserstein distance, \eqref{W1late}. The late-time dynamics actually depends on the choice of $m$. However, if we rewrite \eqref{W1late} to the following form
\begin{align}
    W_1(t) -Z(m) = -\frac{\sqrt{\pi}}{4} \exp\Bigl[-\gamma \Bigl(t- \frac{1}{\gamma}\log m\Bigr)\Bigr] + {\cal O}(e^{-2\gamma t}), 
    \label{shiftW1}
\end{align}
% then the $m$-dependence is actually absorbed into the shift of time. In addition, the redefinition of the 1-Wasserstein distance on the left hand side is the choice of the origin for the measure. In fact, if we measure the distance between $p(t,r)$ and $p(t=\infty,r)$ rather than the original one between $p(t,r)$ and $p(t=0,r)$, it means the left hand side of \eqref{shiftW1}, due to the fact that at $t=\infty$ the state is at $|0\rangle$, and $Z(0)=0$ by our definition \eqref{Zn}.
%
then the $m$-dependence is actually absorbed into the shift of time.
It is convenient to introduce a shifted spatial coordinate
\begin{align}
\widetilde W_1(t) \equiv W_1(t)-Z(m) ,
\label{tildeWdef}
\end{align}
and to redefine $t-\gamma^{-1} \log m$ as the new, shifted time coordinate $t$. 
The quantity~\eqref{tildeWdef} should be regarded as a signed coordinate, whose origin is placed at the
final state $p(t=\infty,r)=p_0(r)$, rather than as a non-negative distance. Indeed, the
actual 1-Wasserstein distance between $p(t,r)$ and the final distribution $p(t=\infty,r)$ is
\begin{align}
W_1\bigl(p(t,r),p(t=\infty,r)\bigr)
= \sum_{k=0}^{m} C_k(t) Z(k) - Z(0)
= Z(m)-W_1(t)
= -\,\widetilde W_1(t) ,
\end{align}
which follows from a calculation similar to Eq.~\eqref{W1W1} and 
%since $p_\infty(r)=p_0(r)$ and
% \begin{align}
% W_1(p_k,p_0)=Z(k)-Z(0)=Z(k) ,
% \end{align}
%where
Eq.~\eqref{W1tzm} and $Z(0)=0$ following from Eq.~\eqref{Zn}.
We mainly use this shifted coordinate $\widetilde W_1(t)$ henceforth, and simply express it as $W_1(t)$ for brevity.

Using the shifted $W_1(t)$, we re-plot the same set of curves given in
%left panel of Fig.~\ref{fig:time2}
Figure~\ref{figure6a}, which results in
%See the right panel of Fig.~\ref{fig:time2}.
Figure~\ref{figure6b}.
Interestingly, all the curves almost overlap with each other, although the analytic functional form of each curve is different from the others. We can argue the bulk picture of these falling particles as follows: the particle starts to fall with some speed at some position in the bulk (these depend on $m$), while it reaches soon a null geodesic due to the acceleration, thus all the trajectories approach the null ray. Thus, the enveloping curve found in the right panel of Fig.~\ref{fig:time2} can be thought of as the common null ray.

What is the functional property of this enveloping curve? At late times, as we have studied in \eqref{shiftW1}, the curve is identical to $-(\sqrt{\pi}/4) \exp[-\gamma t]$.
Let us see the early time behavior. Notice that all the starting points are on the curve. The starting points in the shifted coordinates are
\begin{align}
    (t,W_1) = \left(-\frac{1}{\gamma} \log m, -Z(m)\right).
\end{align}
For large $m$, with the use of the asymptotic form of $Z(m)$ given in \eqref{znsqrtn}, all of these starting points are on a single curve\footnote{%
In fact, the coincidence at early times is not limited to the starting points.  
Using \eqref{W1tzm} together with the explicit coefficients \eqref{Cksol}, the shifted quantity is written as
\begin{equation*}
W_1(t)-Z(m) = - \sum_{k=0}^{m} C_k(t)\, Z(k) .
\end{equation*}
For fixed $t$, the coefficients $C_k(t)$ form a binomial distribution in $k$ (Eq.~\eqref{Cksol}), and then the sum is dominated by values of $k$ around
$k \sim m e^{-\gamma t}$ for large $m$ and at early times, and the shifted distance is approximately given by
\begin{equation*}
W_1(t)-Z(m) \approx - Z(m e^{-\gamma t}) .
\end{equation*}
Then, regarding $t-\gamma^{-1}\log m$ as the shifted time coordinate $t$
as we did around Eq.~\eqref{tildeWdef}, 
this becomes
\begin{equation*}
W_1(t)-Z(m) \approx - Z(e^{-\gamma t}) ,
\end{equation*}
which is independent of $m$. Thus, the approximate overlap of the curves in the right panel
of Fig.~\ref{fig:time2} in early time can be understood not only from the locus of the starting points, but also from the
early-time evolution itself. The asymptotic form \eqref{W1_early} then follows from $Z(n)\sim \sqrt{n}$
for large $n$ (see Eq.~\eqref{znsqrtn}).}
\begin{align}
    W_1 (t) = -\sqrt{\exp(-\gamma t)} = -e^{-\gamma t/2}.
        \label{W1_early}
\end{align}
So we conclude that the common curve behaves as $-e^{-\gamma t}$ for late time and as $-e^{-\gamma t/2}$ at early times (large negative $t$).

To obtain the metric of the geometry, we can follow the strategy described in the previous subsection. 
As the exponential behavior corresponds to the redshift of the geometry, we find that the resultant geometry is
\begin{align}
    ds^2 = \sigma(W_1) \left[
    -f(W_1) dt^2 + \frac{1}{f(W_1)} dW_1^2
    \right]
    \label{W1bh2}
\end{align}
with a blackening function whose asymptotic behavior is given by
(as we can see from $f\bigl(W_1(t)\bigr)= d W_1(t) / dt$ that holds for a null trajectory)
\begin{align}
    f(W_1) 
    \sim \left\{
\begin{array}{cc}
     -\gamma W_1 \qquad &(W_1\sim 0) \\
     -\frac{\gamma}{2} W_1 \qquad &(W_1 \ll -1 )
\end{array}
\right.
\end{align}
Note that both asymptotics are black-hole-like, but with a different surface gravity. In other words, the emergent geometry is Rindler-like, while the acceleration depends on the location: the IR acceleration is bigger by a factor of two compared to the UV acceleration.\footnote{One might wonder why the metric obtained here for the region far from the horizon looks different from that obtained for a fixed trajectory, \eqref{nearbm}.
The reason is that the location of the boundary in the $W_1$ coordinate is different. Here in this section the boundary is at $W_1=-\infty$, while in the previous section the boundary is at $z=0$ (that corresponds to some finite value of $W_1$ in this subsection; the metric \eqref{nearbm} can be identified with Eq.~\eqref{W1bh2} at the locus $W_1 = \sqrt{m}$). The boundary should be properly defined by the behavior of the conformal factor, and here we put some assumption on the location of the boundary.}

\section{Emergent Wasserstein spacetime of SYK}
\label{sec:SYK}

In previous sections we have studied the harmonic oscillator as the simplest setting to test our idea of the emergence of spacetimes through the optimal transport. Here in this section, we study the SYK model \cite{sachdev1992gapless,kitaev2015} which is the most popular quantum mechanical model possessing a dual emergent geometry of AdS${}_2$.
We adopt the idea of Sec.~\ref{sec:HO2}, and will find that the 1-Wasserstein emergent spacetime from the SYK model is consistent with the AdS${}_2$ black hole. This serves as a check that our idea of the optimal transport being a holography in a broad sense actually accommodates the known example of the AdS/CFT correspondence.

The SYK model is a quantum mechanical model made of $N$ Majorana fermions with a disorder average. 
It is by now well established that the low-energy sector of the SYK model is holographically dual to Jackiw–Teitelboim gravity in nearly AdS${}_2$ spacetime \cite{maldacena2016remarks,maldacena2016conformal}.
The strategy we have used in the previous section is to use a Lindbladian system in which a subsystem is coupled to a bath. In the SYK model, the bulk particle-like excitation can be made by an operator quench of a few Majorana operators among $N$, thus we better take a subsystem of a few Majorana fermion, and regard the remaining Majorana fermions as the bath.\footnote{For a general Lidbladian treatment of AdS/CFT, see \cite{Ishii:2025qpy}. Lindbladian analysis of the SYK model was initiated in \cite{kulkarni2022lindbladian}, but it is not about the subsystem of SYK, rather looks at the whole SYK as a subsystem. }

Concretely speaking, in order to make a consistent and minimum Hilbert space of the subsystem, among $N$ Majorana fermions we take two of them, $\chi_1$ and $\chi_2$, to make a creation-annihilation operator
\begin{align}
    \hat{c}\equiv \hat{\chi}_1 + i \hat{\chi}_2, \quad 
    \hat{c}^\dagger = \hat{\chi}_1 - i \hat{\chi}_2.
\end{align}
The Hilbert space of this subsystem is a qubit, spanned by $|0\rangle$ and $|1\rangle$, satisfying 
\begin{align}
    \hat{c}\ket{0} = 0, \quad \hat{c}\ket{1}=\ket{0},
    \nonumber \\
        \hat{c}^\dagger\ket{0} = \ket{1}, \quad \hat{c}^\dagger\ket{1}=0.
\end{align}
The commutation relation is $\{\hat{c},\hat{c}^\dagger\}=1$, and the fermion number operator $\hat{n}\equiv \hat{c}^\dagger \hat{c}$ shows $\hat{n}\ket{n}=n\ket{n}$.

Let us build the Lindbladian of this subsystem. The original SYK interaction is of the form $J_{ijkl}\chi_i\chi_j\chi_k\chi_l$, thus dividing the system into the subsystem and the bath, we obtain three possibilities;
\begin{align}
    \chi_I\chi_J\chi_K\chi_L, \quad \chi_{1,2}\chi_I\chi_J\chi_K, \quad
    \chi_1 \chi_2 \chi_I\chi_J,
\end{align}
where $I,J,K=3,4,\cdots, N$ are the bath fermion labels. This means that in the Lindbladian the jump operators are 
\begin{align}
    \hat{L}_1 \equiv \hat{c}, \, \hat{L}_2 \equiv \hat{c}^\dagger, \, \hat{L}_3 \equiv \hat{c}^\dagger \hat{c}.
\end{align}
Using these, we can define the Lindblad master equation
\begin{align}
    \partial_t \hat{\rho} = -i [\hat{H}_0, \rho]
    + \sum_i \gamma_i \left(
    \hat{L}_i \hat{\rho}\hat{L}_i^\dagger - \frac12 \{ \hat{L}_i^\dagger \hat{L}_i, \hat{\rho}\}
    \right),
    \label{master}
\end{align}
where the coupling $\gamma_i$ ($i=1,2,3$) is positive.
For the subsystem Hamiltonian $\hat{H}_0$, we expect $\hat{H}_0=0$ at low energy because the SYK model at low energy is scale-independent and has no mass gap. So hereafter we put $\hat{H}_0=0$.

Since the Hilbert space is just two-dimensional, we parametrize the density matrix as
\begin{align}
    \hat{\rho}(t) = \left(
    \begin{array}{cc}
         \rho_{11}(t)&\rho_{10}(t)  \\
         \rho_{01}(t)& \rho_{00} (t)
    \end{array}
    \right).
\end{align}
The master equation \eqref{master} in this parametrization reduces to
\begin{align}
    \partial_t \rho_{11} & = -\gamma_1 \rho_{11} + \gamma_2 \rho_{00},
    \\
    \partial_t \rho_{00} & = -\gamma_2 \rho_{00} + \gamma_1 \rho_{11},
    \\
        \partial_t \rho_{10} & = -\frac12\left(
        \gamma_1 + \gamma_2 +\gamma_3\right) \rho_{10},
\\
        \partial_t \rho_{01} & = -\frac12\left(
        \gamma_1 + \gamma_2 +\gamma_3\right) \rho_{01}.
\end{align}
From these equations, first we find that the off-diagonal components $\rho_{10}$ and $\rho_{01}$ are decoupled from the diagonal components, and they just decay. This is the decoherence effect and we just consider completely decohered states, by putting $\rho_{01}=\rho_{10}=0$.
Next, we look at the diagonal components. As there is no directed orientation to prefer the state $\ket{0}$ or $\ket{1}$, we can simply assume $\gamma_1=\gamma_2 (\equiv \gamma)$. We choose a decohered initial condition
\begin{align}
    \hat{\rho}(t=0) = \left(
    \begin{array}{cc}
         \rho_{11}(t=0)&0  \\
         0&1-\rho_{11}(t=0)
    \end{array}
    \right),
\end{align}
then the solution to the Lindblad master equation is
\begin{align}
    \rho_{11}(t) = \left(\rho_{11}(t=0)-\frac12\right)e^{-2\gamma t} + \frac12 , \\
    \rho_{00}(t) = \left(\frac12-\rho_{11}(t=0)\right)e^{-2\gamma t} + \frac12 .
\end{align}
Restricting to the decohered diagonal states, we identify the diagonal components of the density matrix, $(\rho_{11}(t), \rho_{00}(t))$, with a probability distribution on a two-point discrete space.
Then, denoting the Wasserstein distance between the density matrices diag$(1,0)$ and diag$(0,1)$ as $d$, we find the time evolution of the Wasserstein distance as
\begin{align}
    W_p = d\, \Bigl| \rho_{11}(t=0)-\frac12 \Bigr|^{1/p} \left(1-e^{-2\gamma t}\right)^{1/p}.
\end{align}
At low energy, the only scale of the SYK model is the temperature $T$, so we expect $\gamma = c T/2$ with some positive numerical constant $c$. Resultantly, the time evolution of the $p$-Wasserstein distance in this subsystem SYK is
\begin{align}
    W_p \propto \bigl(1-e^{-c T t}\bigr)^{1/p}.
    \label{sykWp}
\end{align}
In particular, we are interested in $1$-Wasserstein distance, which is given by
\begin{align}
    W_1 \propto 1-e^{-c T t}.
    \label{sykW1}
\end{align}

In the following, we will see that the Wasserstein distance \eqref{sykW1} is consistent with a particle falling in the AdS${}_2$ Schwarzschild black hole geometry\footnote{This form \eqref{AdS2sch} of the metric may not be of the popular form of the AdS${}_2$ Schwarzschild
$$ds^2 = -(r^2-r_*^2)dt^2 + \frac{1}{r^2-r_*^2}dr^2$$ but one can show that they are equivalent through the coordinate transformation $z/z_* \equiv 2r_*/(r+r_*)$, with $z_*\equiv 1/(2r_*)$. This form \eqref{AdS2sch} is shared with the popular metric form of higher-dimensional AdS${}_{d+1}$ black holes, with $d=1$.}
\begin{align}
    ds^2 = \frac{1}{z^2} \left(
    -\frac{z_*-z}{z_*} dt^2 + \frac{z_*}{z_*-z}dz^2
    \right).
    \label{AdS2sch}
\end{align}
Here, the horizon is at $z=z_*$, and the relation to the temperature is $z_* = 1/(4\pi T)$. 
We identify the Wasserstein distance as the radial coordinate of the emergent spacetime, $W_p=z$. 
Following the same logic that we used in the previous section, we go to a conformally flat spacetime
\begin{align}
    ds^2 = \frac{1}{z_*^2}
    \frac{e^{-s/z_*}}{(1-e^{-s/z_*})^2} (-dt^2+dy^2),
\end{align}
by a coordinate transformation 
\begin{align}
    z = z_* (1-e^{-y/z_*}).
\end{align}
Then using the relation \eqref{ident}, 
we find that the relation between the radial location of the particle at $z=W_1$
at time $t$ of the AdS${}_2$ boundary is
\begin{align}
    W_1 \propto 1-e^{-4\pi T t}.
\end{align}
This has the same form as the Wasserstein time evolution of the SYK subsystem, \eqref{sykW1}.

%%%%%%%%%%%%%%%%%%%%%%%%%%%%%%%%%%%%%%%%%%%%%%%%%%%%%%%%%%%%%%%%

\section{Wasserstein distance is a generalized Krylov complexity}
\label{sec:WK}

\subsection{Wasserstein and Krylov: the relationship}

Looking back the formula of the time evolution of the 1-Wasserstein distance \eqref{W1W1} for the Lindblad harmonic oscillator
\begin{align}
    W_1(t) = \sum_{k=0}^m C_k(t) W_1(p_k,p_m)
    \label{W1tW1}
\end{align}
with the density matrix given by \eqref{dent}, 
it has a strong similarity with the formula for a Krylov complexity:
\begin{align}
    {\cal K}(t) = \sum_{n=0}^\infty 
    n \left|\phi_n(t)\right|^2,
    %\qquad \mbox{with}\qquad C_n(t) \equiv |\phi_n(t)|^2.
    \label{KC}
\end{align}
if we regard $C_k(t)$ and $W_1(p_k,p_m)$ as the counterparts of $\left|\phi_n(t)\right|^2$ and $n$, respectively.
In this section, we argue that our 1-Wasserstein distance can be thought of as a generalized Krylov complexity.

Let us remind the readers of how the Krylov complexity for an operator is defined. For a given quantum mechanical system with a Hamiltonian and for a given operator ${\cal O}$, one defines a Krylov basis $\{{\cal O}_n\}$ by subsequently acting the Liouvillian $[H,*]$ on the operator $n$ times while making the produced operators to be orthogonal to each other: ${\rm Tr}[{\cal O}_n {\cal O}_m] = \delta_{nm}$. Then one expands the Heisenberg time-evolved operator ${\cal O}(t) \equiv e^{iHt}{\cal O}e^{-iHt}$ in terms of this Krylov basis,
\begin{align}
    {\cal O}(t) = \sum_n \phi_n(t) {\cal O}_n,
    \label{HOH}
\end{align}
with ${\cal O}_0 = {\cal O}$. Using this basis expansion, the spreading of the operator in time in the operator space of the given quantum system is defined by the Krylov complexity, \eqref{KC}. This definition was generalized to open systems with the Lindblad formulation \cite{bhattacharya2022operator,liu2023krylov,bhattacharjee2023operator,bhattacharya2023krylov,bhattacharjee2024operator}, in which case the time-evolved operator ${\cal O}(t)$ is defined through the Lindblad master equation.

In our case, obviously the operator in question is the density matrix. Let us consider the operator ${\cal O}=\hat{\rho}^{1/2}$. The reason why this definition needs a square root is to satisfy the normalization condition of the Krylov operator, ${\rm Tr}[{\cal O}^2] = {\rm Tr}[\hat{\rho}]=1$. The Lindbladian used in Sec.~\ref{sec:HO2} is with the jump operator being the annihilation operator $\hat{a}$, meaning that, when the initial density operator is chosen as $\ket{m}\bra{m}$, the Krylov basis is given by the Gram-Schmidt orthogonalization as
\begin{align}
    {\cal O}_0 = \ket{m}\bra{m}, \quad
    {\cal O}_1 = \ket{m-1}\bra{m-1}, \quad
    {\cal O}_2 = \ket{m-2}\bra{m-2}, \quad \cdots.
\end{align}
This is indeed the basis with which we expand the density matrix~\eqref{dent},
and then $\phi_n(t)$ in Eq.~\eqref{HOH} for ${\cal O}=\hat\rho^{1/2}$ is identified with $C_n^{1/2}$.
Thus, except that in our Wasserstein distance \eqref{W1tW1} we use $W_1(p_k, p_m)$ instead of $n$ in the Krylov complexity, the two formulas \eqref{W1tW1} and \eqref{KC} are the same.
The 1-Wasserstein distance is given by $W_1(p_k,p_m) = Z(m)-Z(k)$, while in the corresponding Krylov complexity one has $n=m-k$, which counts the number of the annihilation operators to reach the vector 
${\cal O}_n = \ket{k}\bra{k}$ starting from the original ${\cal O}_0 = \ket{m}\bra{m}$. 

This difference, the measure $Z(n)$ rather than $n$, is in the realm of ``generalized Krylov complexity" \cite{fu2025statistics, fan2023generalised, camargo2025higher} in which the ``$n$" in \eqref{KC} is replaced by its arbitrary power $n^\delta$. Therefore, we conclude that the 1-Wasserstein distance, as a cost for the optimal transport, is a generalized Krylov complexity. 

Here we emphasize the necessary condition that the 1-Wasserstein distance \eqref{W1tW1} becomes a generalized Krylov complexity. Remember that in the derivation of \eqref{W1tW1} or \eqref{W1W1}, we have used the fact that the CDFs are ordered, {\it i.e.} the CDFs at different times do not intersect with each other, such that in the definition of the Wasserstein distance the absolute value of the difference of the CDFs is replaced by just the difference. This means that the similarity between the 1-Wasserstein distance and the generalized Krylov complexity is found only when the CDFs are ordered. 

Interestingly, 
we can find the ``Wasserstein operator" which, for the 1-Wasserstein distance, serves as a complexity operator for the generalized Krylov complexity, as follows. The complexity operator $\hat{\cal K}$ is the operator whose expectation value with the Schr\"odinger wave function provides the state spread complexity, $K(t) = \bra{\psi(t)}\hat{\cal K}\ket{\psi(t)}$ \cite{balasubramanian2022quantum}. We will find this complexity operator for our Wasserstein distance.
Remember that we are measuring the 1-Wasserstein distance for the Husimi Q-representation whose CDF is
\begin{align}
    F(|\alpha|,t)= \int_0^{|\alpha|} \! d|\beta|\,  2\pi |\beta| Q(|\beta|,t),
\end{align}
for the states whose Husimi representation is rotationally symmetric.\footnote{Any rotationally symmetric Husimi Q-function of a trace-class density matrix (which automatically provides an even function of $|\alpha|$) can be written as a linear combination of $Q_n(|\alpha|)$. Thus, that corresponds to a Husimi Q-representation of $\rho=\sum_n c_n \ket{n}\!\bra{n}$ for a real normalized vector $c_n$.} Assuming that the CDFs are ordered, we can rewrite the 1-Wasserstein distance by a partial integration as
\begin{align}
    W_1(t) &\equiv \int_0^\infty \bigl(F(|\alpha|,0)-F(|\alpha|,t)\Bigr)
    \\
    &=\int_0^\infty d|\alpha| \, 2\pi|\alpha|^2\bigl(Q(|\alpha|,t)-Q(|\alpha|,0)\Bigr).
\end{align}
So, defining the coordinate 
\begin{align}
    Z_1(t) \equiv \int_0^\infty d|\alpha| \, 2\pi|\alpha|^2 Q(|\alpha|,t),
\end{align}
we find $W_1(t) = Z_1(t) - Z_1(0)$. This coordinate is written as an expectation value,
\begin{align}
    Z_1(t) & =\int_0^\infty d|\alpha| \, 2|\alpha|^2 \braket{\psi(t)}{\alpha} \braket{\alpha}{\psi(t)}  \\
    & = \bra{\psi(t)} \hat{\cal W}\ket{\psi(t)}
\end{align}
where the ``Wasserstein operator" $\hat{\cal W}$, which serves as a complexity operator, is defined as
\begin{align}
    \hat{\cal W} \equiv \int_0^\infty \! d|\alpha| \, 2|\alpha|^2 \ket{\alpha}\bra{\alpha}.
    \label{defWop}
\end{align}
Therefore we conclude that with this Wasserstein operator, the 1-Wasserstein distance is given just by its quantum expectation value.\footnote{Note that here the expectation value is purely quantum mechanical standard one, while the Kantorovich-Rubinstein duality which we reviewed in \eqref{f*def} is with the expectation value based on the distributions (which are the Husimi Q-representation in our case), so they are a little different from each other. See the footnote \ref{footH} too.} The physical meaning of this expression is important. In fact, the integrand $|\alpha|^2$ is equal to the classical Hamiltonian measured with the coherent state basis $\ket{\alpha}$. Therefore, the Wasserstein operator is measuring the energy.\footnote{Note that the Wasserstein operator \eqref{defWop} is with the radial integration of the complex variable $\alpha$. This works since in our setup the Husimi representation is rotationally symmetric in the phase space. One can rewrite \eqref{defWop} also as $\hat{\cal W} = \int\! \frac{d^2\alpha}{\pi} \, |\alpha| \ket{\alpha}\bra{\alpha}$, which may work for more general situation. With this expression, we may say that the Wasserstein operator measures the square root of the energy.}

\subsection{Unique features of Wasserstein distance compared to Krylov complexity}
\label{sec:unique}

Important unique features of our 1-Wasserstein distance, compared to the generalized Krylov complexity, are as follows.
\begin{itemize}
    \item The distance does not depend on the initial choice of the operator. The 1-Wasserstein distance forms a metric space and can define distance between any distributions, while the Krylov space depends on the choice of the initial operator. Thus, with the 1-Wasserstein distance, the distance (growth) can be compared among different choices of the initial operators.
    \item Once a single basis space is employed, the distance can be measured independently of the time evolution dynamics. A natural choice for the basis space is the set of the energy eigenstates or their decohered (diagonal) density matrix. This natural choice is produced by the creation/annihilation operator as the jump operator of the Lindbladian. In other words, the complexity is counted by the number of the creation/annihilation operator, which follows the standard definition of the quantum computational complexity, where in our case the universal gate set can be provided by the creation/annihilation operator.
\end{itemize}
In the following, to demonstrate these merits of the optimal transport technique, we present two analyses: one is the growth of the density matrix by the creation operator (as opposed to the one in Sec.~\ref{sec:HO2} using the annihilation operator to go down), and the other is the Lindblad time evolution among thermal states. Both of these are using the same 1-Wasserstein space, so the result can be directly and quantitatively compared among these examples.

First, let us consider a hypothetical ``bath" in which the jump operator is the creation operator $\hat{a}^\dagger$, rather than the annihilation operator used in Sec.~\ref{sec:HO2}. With this choice, the state time-evolves upward, rather than going down. The Lindblad master equation is 
\begin{align}
    \frac{d}{dt}\hat{\rho} = - i \omega [\hat{n},\hat{\rho}]
    + \gamma \left(
    \hat{a}^\dagger\hat{\rho}\hat{a} - \frac12 \{ (\hat{n}+1), \hat{\rho}\}
    \right).
    \label{Lind2}
\end{align}
With the Husimi Q-representation, we obtain the equivalent description\footnote{Note that when this is viewed as the Fokker-Planck equation, the dissipation term is missing. }
\begin{align}
    \frac{\partial}{\partial t} p(t,r) = 
    \left[-\frac{\gamma}{2}-\frac{\gamma}{2} r \frac{\partial}{\partial r}\right]p(t,r).
\end{align}
A generic solution to this equation is
\begin{align}
    p(t,r) = \frac{1}{r}f\left(-\gamma t/2 + \log r\right)
\end{align}
with an arbitrary function $f$. 

As an example, let us take the initial quantum state as $\hat{\rho} = \ket{0}\bra{0}$. Then we find
\begin{align}
    p(t,r) = 2 e^{-\gamma t/2}  r(t) e^{-r(t)^2}
\end{align}
where we have defined $r(t)\equiv r e^{-\gamma t/2}$. The 1-Wasserstein distance can be calculated as
\begin{align}
    W_1(t) = \frac{\sqrt{\pi}}{2} \left(e^{\gamma t/2}-1\right).
    \label{W100}
\end{align}
So the density operator spreads exponentially in time.

As another example, let us take the initial quantum state as $\hat{\rho} = \ket{m}\bra{m}$ with $m \gg 1$. In this case, the initial Husimi Q-representation is a Gaussian, \eqref{Gauss}. The time-evolved one can be easily obtained as
\begin{align}
    p(t,r) \sim \sqrt{\frac{2}{\pi}} e^{-\gamma t/2}\exp[-2 \left(r(t)-\sqrt{m+1/2}\right)^2].
    \label{Gauss2}
\end{align}
Using this, the time evolution of the Wasserstein distance is 
\begin{align}
    W_1 = \sqrt{m}\left(e^{\gamma t/2}-1\right) + {\cal O}(1).
\end{align}
Again, this grows exponentially in time.

Next, let us consider the finite temperature bath.
The Lindblad master equation with the bath at thermal equilibrium with temperature $T$ is given by\footnote{The master equation \eqref{Lind} used in Sec.~\ref{sec:HO2} is \eqref{Lindn} with $\bar{n}=0$, that is, a zero temperature bath.}
\begin{align}
    \frac{d}{dt}\hat{\rho} = - i \omega [\hat{n},\hat{\rho}]
    + \gamma (\bar{n}+1)\left(
    \hat{a}\hat{\rho}\hat{a}^\dagger - \frac12 \{ \hat{n}, \hat{\rho}\}
    \right)
    + \gamma \bar{n}\left(
    \hat{a}^\dagger\hat{\rho}\hat{a} - \frac12 \{ (\hat{n}+1), \hat{\rho}\}
    \right).
    \label{Lindn}
\end{align}
Here $\bar{n}$ is the excitation number at the thermal equilibrium,   
\begin{align}
    \bar{n}\equiv \frac{1}{e^{\beta \omega}-1}.
\end{align}
Going to the Husimi Q-representation, we obtain a Fokker-Planck equation
\begin{align}
    \frac{\partial}{\partial t}Q 
    = \frac{\partial}{\partial\alpha}
    \left(\left(\frac{\gamma}{2}+i \omega\right)\alpha Q\right)
    +
    \frac{\partial}{\partial\alpha^*}
    \left(\left(\frac{\gamma}{2}-i \omega\right)\alpha^* Q\right)
    +\gamma(\bar{n}+1)\frac{\partial^2}{\partial\alpha\partial\alpha^*}Q.
\end{align}
As an initial condition, we choose a thermal state with a different value of the temperature,
\begin{align}
    \hat{\rho}(t=0)
    = (1-e^{-\tilde{\beta}\omega})\sum_{n=0}^\infty e^{- n \tilde{\beta} \omega} \ket{n}\bra{n}.
\end{align}
With this state, the expectation value of the number operator is
\begin{align}
    \tilde{n} \equiv 
    \frac{1}{e^{\tilde \beta\omega}-1}.
\end{align}
We can solve the Fokker-Planck equation, and the solution is
\begin{align}
    Q(t) = \frac{g(t)}{\pi}\exp\left[-g(t) \alpha\alpha^*\right],
    \label{QtFP}
\end{align}
where
\begin{align}
    g(t)\equiv \left[(\bar{n}+1)(1-e^{-\gamma t}) + (\tilde{n}+1)e^{-\gamma t}\right]^{-1}.
\end{align}
That is, the Gaussian distribution centered at the origin $\alpha=0$ with the width evolving as $g(t)^{-1/2}$ is the solution of the Fokker-Planck equation.

Using the general formula for the $p$-Wasserstein distance between Gaussian distributions, we find that the time evolution, or the distance between any states at time $t_1$ and time $t_2$, is given by
\begin{align}
    W_p(t_1, t_2) = \Bigg|
    \frac{1}{\sqrt{g(t_1)}} - \frac{1}{\sqrt{g(t_2)}}
    \Bigg|
    \left(\Gamma(1+p/2)\right)^{1/p}.
    \label{thermW1}
\end{align}
This $W_p$ is a nice space which allows the additive distance rule as in the case of $W_1$, since for this special choice of the initial states the $p$-Wasserstein distance allows the same structure as that of $W_1$, and the only difference is the normalization coefficient (which is the last factor in \eqref{thermW1}).
The time evolution shows that the local temperature of the subsystem harmonic oscillator will evolve and finally approach the bath temperature with the time scale governed by the decay rate $\gamma$.\footnote{
The solution $Q(t)$ stays within the family of thermal Gaussian distributions (i.e., the Husimi $Q$-representation of thermal states). Indeed, writing Eq.~\eqref{QtFP} as
\[
Q(t)=\frac{g(t)}{\pi} e^{-g(t)|\alpha|^2}
=\frac{1}{\pi(n(t)+1)}
\exp\!\left[-\frac{|\alpha|^2}{n(t)+1}\right],
\]
we identify the time-dependent expectation value of the number operator as
\[
n(t)=\frac{1}{g(t)}-1
=\bar n+(\tilde n-\bar n)e^{-\gamma t}.
\]
Hence, the state at each time can be regarded as a thermal state with an effective inverse temperature $\beta_{\rm eff}(t)$ defined through
\[
n(t)=\frac{1}{e^{\beta_{\rm eff}(t)\omega}-1}\,,
\qquad
\beta_{\rm eff}(t) = \frac1\omega \log\left( 1 + \frac{1}{n(t)}\right)
\]
This shows that the subsystem relaxes to the bath equilibrium, $\beta_{\rm eff}(t)\to\beta$ as $t\to\infty$. 
%While $\beta_{\rm eff}(t)$ is a nonlinear function of $n(t)$, 
The occupation number itself relaxes exactly exponentially with rate $\gamma$, and hence $\beta_{\rm eff}(t)$ approaches $\beta$ asymptotically with the same rate.}
The Wasserstein distance evolution shares the same character. 

Now, as we advertised, for our case of the Wasserstein distances, we can compare the growth directly between the first one (with the creation operator as the jump operator) and the second one (with the thermal evolution), because the basis is shared.
The first one grows exponentially in time, while the second one decays exponentially. Even for a high temperature $\bar{n}\gg 1$ two systems do not coincide. Therefore
we see, from the difference in the growth of the Wasserstein distances, that the Lindbladian with the creation operator as its jump operator is quite different from the Lindbladian with the high bath temperature.

\section{Summary and discussion}
\label{sec:sum}

In this paper, we applied the notion of optimal transport and the manifold hypothesis to a harmonic oscillator and the SYK model to find that the Wasserstein space can be regarded as a holographic spacetime.

In Sec.~\ref{sec:HO1}, using the Euclidean embedding analysis motivated by the manifold hypothesis, we have shown that for the Wasserstein space to be exactly one-dimensional for the harmonic oscillator energy eigenstates, we need to choose the 1-Wasserstein distance for the Husimi Q-representation of the states. The Wasserstein space admits a metric \eqref{metgzz}. 

In Sec.~\ref{sec:HO2}, we employed a Lindblad harmonic oscillator to find the 1-Wasserstein time evolution (Fig.~\ref{fig:W1t}). Identifying the Wasserstein distance as the radial coordinate of the holographic spacetime, a particle trajectory falling down in the holographic geometry
is reproduced by the Wasserstein time evolution, once the metric is chosen as that of a black hole, \eqref{W1bh}. Different trajectories with different initial states are found to be shared, and the set of those trajectories consistently produces a single Wasserstein spacetime \eqref{W1bh2}. 

In Sec.~\ref{sec:SYK}, we consider a two-Majorana-fermion subsystem of SYK and the Lindblad analysis provides a $p$-Wasserstein time evolution \eqref{sykWp}. The time evolution is consistent with a null shock wave in AdS${}_2$ Schwarzschild geometry observed from the boundary, when the radial coordinate is identified with the $p$-Wasserstein distance
for $p=1$.

In Sec.~\ref{sec:WK}, we have shown that our 1-Wasserstein distance \eqref{W1tW1} in the harmonic oscillator case coincides with a generalized Krylov complexity. Furthermore, we pointed out that the 1-Wasserstein distance provides a more universal metric space which is independent of the initial state, by looking at the examples of the Lindbladian with the jump operator being a creation operator \eqref{W100}, and with the generic thermal behavior of time evolution of the thermal states \eqref{thermW1}.

The results obtained in the last two sections are encouraging, in the sense that our whole philosophy of treating the holographic principle as a Wasserstein space with the manifold hypothesis seems to be working nicely. However, our demonstration in Sec.~\ref{sec:HO1} and Sec.~\ref{sec:HO2} is just for a harmonic oscillator; thus, a more thorough analysis using holographic CFT would be necessary to check if our philosophy really works as a fundamental view of the holographic principle or not. There are various future directions of research concerning this, and we hope to report on some in the near future.

We have several points for discussion to close this paper, in particular on (i) the possible distance measure, and (ii) the representation of the distributions. 
\begin{description}
    \item[(i) Comparison in the distance measures.] 
In this paper, we have employed the $p$-Wasserstein distances as the distance measures in the distribution space of quantum mechanics. The main reason is, as described in Sec.~\ref{sec:intro}, the holographic space should describe the optimal path of a probe particle in the bulk. Here we describe another reason.
Although the Hilbert space distance measures such as Fubini-Study distance, trace distance and Bures distance have been widely used in the context of holography, we here employ optimal transport distances, simply because the latter can distinguish the distance between basis vectors that are mutually orthogonal. Consider energy eigenstates; they form an orthogonal basis, thus the Hilbert space distance measure based on the norm and the inner product provides only degenerate and vanishing distance against all orthogonal states, making it difficult to capture the whole geometric structure of the Hilbert space.

In addition, in Sec.~\ref{sec:WK} we pointed out that our Wasserstein distance is regarded as a generalized Krylov complexity. In fact, for the holographic complexity, there have been some early discussions on whether it can be described by the Hilbert space norm or not \cite{hashimoto2018thoughts,yang2019unitary, ali2019time}. The norm is unitary invariant, which means that it is independent of the basis. On the other hand, the Wasserstein distance is not unitary invariant, thus basis-dependent. Thus, our result points to the direction that the unitary invariance of the measures could be a too-strong condition for obtaining a defining distance in holography from quantum mechanical distributions.

Wasserstein distances can be compared with the Kullback-Leibler divergence (or the cross entropy). The latter is widely used in AI and machine learning, while it is not appropriate for our purpose, due to the same reason as for the Hilbert space norm: the Kullback-Leibler divergence diverges among the distributions that share no common support, meaning that orthogonal quantum states cannot be distinguished. On the other hand, the optimal transport distances can distinguish them with a finite expression. Therefore, the optimal transport is not only in accord with the bulk picture of a moving probe particle but also serves technically as a better distance measure that can distinguish quantum states.

\item[(ii) Other choice of the representation.] 
We have discarded the probability distribution $\rho(x)$ as our choice of the representation because of the incapability in the embedding dimensions in Sec.~\ref{sec:HO1}, but the suitability actually depends on the choice of the quantum states. Rather than the energy eigenstates we employed in this paper, with some other choice of quantum states the probability distribution could play the central role. As an example, let us choose instead a set of coherent states $\ket{\alpha}$ with $\alpha = e^{i\theta}\alpha_0$ with a constant real number $\alpha_0$ and a variable $\theta$ $(0\leq \theta < 2\pi)$, as the quantum states of the reference. Then for $\rho(x,\theta) \equiv |\langle x|\alpha\rangle|^2$ we find $W_p(\theta_1, \theta_2)\propto |\cos\theta_1 - \cos\theta_2|$, with which a well-defined one-dimensional space is emergent. Furthermore, the $\theta$ space is nothing but the space in which the time evolution follows. Thus, in this example, just the ordinary probability distribution may have a nice holographic property. The difference from and the relevance to the Husimi Q-representation should be studied in future work.
\end{description}

Finally, we comment on recent developments in quantum Wasserstein distances. As reviewed in \cite{beatty2025wasserstein}, there are various versions of Wasserstein distances through optimal transport in the space of density matrices. On the other hand, the Wasserstein distance used in this paper is the optimal transport between Husimi Q-representation of quantum states \cite{zyczkowski1998monge}, not the direct distance between quantum density matrices themselves. So, how our analyses can be generalized with the use of the purely quantum Wasserstein distances is of interest. In fact, in \cite{de2021quantum} it was shown that the Husimi Wasserstein distance provides a certain lower bound for a quantum Wasserstein distance. Unfortunately, 
some versions of the quantum Wasserstein distances do not satisfy the triangle inequality, which is the main reason why we are not so much motivated to study the quantum Wasserstein distances. However, since the quantum Wasserstein distances are expected to be directly related to quantum operations, we hope that holographic operations in the bulk side will finally be interpreted as some quantum operations on the quantum Wasserstein distances. This is also left to some future work.

%%%%%%%%%%%%%%%%%%%%%%%%%%%%%%%%%

\section*{Acknowledgments}

We would like to thank S.~Ito for valuable discussions and his lectures on optimal transport. We also thank T.~Furugori, K.~Kyo, S.~Sugishita, T.~Yoda and S.~Yoshikawa for insightful discussions.
The work of K.~H.~was supported in part by JSPS KAKENHI Grant No.~JP22H01217, JP22H05111 and JP22H05115.
The work of N.~T.~was supported in part by JSPS KAKENHI Grant No.~JP21H05189, JP22H05111 and 25K07282.

\appendix
\section{Numerical results for the embedding dimensions}
\label{appA}

%This appendix provides supplementary numerical data to support the analysis in Sec.~\ref{sec:2.2}.

This appendix provides supplementary numerical data to support the analysis in Sec.~\ref{sec:2.2}. The arguments therein were based on the numerical results for $N=6$ and $p=1,\ldots, 4$.
We show numerical evidence that qualitative features of those results persist for general $N$ and $p$ in this appendix.

\subsection{Probability distribution}
\label{appA:prob}

We first summarize supplementary numerical checks for the probability distribution $\rho_n(x)$ in the position space. In Sec.~\ref{sec:2.2}, the analysis of the embedding dimensions was based on the distance matrices for $N=6$ and $p=1,\ldots,4$. In particular, Table~\ref{tab:ev1} shows that for $N=6$ the Gram matrix has a negative eigenvalue only for $p=1$, while for $p=2,3,4$ all non-zero eigenvalues are positive. Here we report additional numerical checks for larger $N$, in order to see whether this pattern persists.

Our additional numerical checks indicate that the $p=2$ case is qualitatively different from the cases with $p>2$. Indeed, in one dimension the squared $2$-Wasserstein distance is represented as the squared $L^2$ distance between quantile functions, so the corresponding Gram matrix is analytically guaranteed to be positive-semidefinite.

For $p\geq 3$, by contrast, the absence of negative eigenvalues at $N=6$ turns out to be only a finite-$N$ phenomenon. By increasing $N$, negative eigenvalues of the Gram matrix do appear, and their onset moves to smaller $N$ as $p$ increases. Numerically, the onset is relatively late for $p=3$, around $N\sim 40$, while for $p=4$ it is already visible around $N\sim 12$. For still larger values, the tendency becomes stronger: in our scan the onset is again around $N\sim 12$ for $p=5$, and already around $N\sim 10$ for $p=6$. We also observe that, at a fixed sufficiently large $N$, the number of negative eigenvalues tends to increase as $p$ is increased. In this sense, increasing $p$ makes the non-Euclidean character of the probability-distribution distance more pronounced.

This means that the values of $D_{\min}$ and $D_{\mathrm{eff}}$ quoted in Table~\ref{tab:ev1} for the probability distribution at $N=6$ should also be interpreted as a small-$N$ feature for $p\geq 3$. Once a negative eigenvalue appears, the Euclidean embedding itself becomes impossible, and $D_{\min}$ and $D_{\mathrm{eff}}$ are no longer defined. By contrast, for $p=2$ the positive-semidefinite nature of the Gram matrix is guaranteed analytically, so this obstruction does not arise.

\subsection{Husimi Q-distribution}
\label{appA:hq}

Next, for the Husimi Q-distribution,
we show the result for a larger system size $N=16$ here as an example, while we confirmed that the results are qualitatively similar for different $N$.

The $p$-Wasserstein distances for the Husimi Q-distribution are evaluated using the quantile function $F_{p_n}^{-1}$,
whose inverse (cumulative distribution function $F_{p_n}$) is expressed via the regularized incomplete gamma function:
\begin{equation}
    F_{p_n}(r) 
    := \int_0^r p_n(r)
    = 1 - \frac{\Gamma(n+1,\,r^2)}{\Gamma(n+1)}\,,
\end{equation}
where $p_n(r)$ is defined by Eq.~\eqref{HQp}.
Then, the quantile function ($F_{p_n}^{-1}$) can be expressed as
\begin{equation}
    F_{p_n}^{-1}(f) 
    = \sqrt{q^{-1}(n+1,1-f)}
    \,,
\end{equation}
where $q^{-1}(n+1,1-f)=r^2$ is the inverse of the regularized Gamma function satisfying $1-f = \Gamma(n+1,r^2)/\Gamma(n+1):=q(n+1,r^2)$.
The function $q^{-1}$ is implemented in Mathematica as \texttt{InverseGammaRegularized}, and it can be evaluated with arbitrary precision.

Table~\ref{tab:evN16} shows the leading and trailing eigenvalues of the Gram matrix $B$
for $N=16$ states and $p=1,\dots,6$, confirming the three features observed in Sec.~\ref{sec:2.2}
at larger $N$.

\begin{table}[h]
    \centering
    \begin{tabular}{c|ccc c cc|c|c}
     & $\lambda_1$ & $\lambda_2$ & $\lambda_3$
     & $\cdots$ & $\lambda_{15}$ & $\lambda_{16}$
     & $D_{\min}$ & $D_{\mathrm{eff}}$ \\
    \hline
    $p=1$ &
      12.9893 & $0$ & $0$ & $\cdots$ & $0$ & $0$
      & 1 & 1 \\
    $p=2$ &
      12.9936 & $1.4\!\times\!10^{-3}$ & $3.6\!\times\!10^{-6}$
      & $\cdots$ & $4.2\!\times\!10^{-38}$ & $0$
      & $N-1$ & 1.000209 \\
    $p=3$ &
      12.9977 & $2.5\!\times\!10^{-3}$ & $-8.5\!\times\!10^{-6}$
      & $\cdots$ & $9.9\!\times\!10^{-37}$ & $0$
      & -- & -- \\
    $p=4$ &
      13.0016 & $3.4\!\times\!10^{-3}$ & $-2.8\!\times\!10^{-5}$
      & $\cdots$ & $-5.7\!\times\!10^{-36}$ & $0$
      & -- & -- \\
    $p=5$ &
      13.0055 & $4.1\!\times\!10^{-3}$ & $-5.1\!\times\!10^{-5}$
      & $\cdots$ & $-2.3\!\times\!10^{-37}$ & $0$
      & -- & -- \\
    $p=6$ &
      13.0091 & $4.7\!\times\!10^{-3}$ & $-7.4\!\times\!10^{-5}$
      & $\cdots$ & $1.9\!\times\!10^{-35}$ & $0$
      & -- & -- \\
    \end{tabular}
    \caption{Leading and trailing eigenvalues $\lambda_k$ and embedding
    dimensions for $W_p$ of the Husimi Q-distribution with $N=16$ states.
    %,    computed at \texttt{WorkingPrecision}~$=60$.
    For $p=1$, $\lambda_{k>1}=0$ exactly.
    For $p\geq 3$, $\lambda_3<0$ so no Euclidean embedding exists
    and $D_{\min}$, $D_{\mathrm{eff}}$ are undefined (``--'').
    The structural zero $\lambda_{16}=0$ is required by $\det J=0$.}
    \label{tab:evN16}
\end{table}

\begin{enumerate}

\item[(i)] \textbf{$D_{\min}=N-1$ for $p=2$; $D_{\min}=1$ for $p=1$.}
For $p=1$, only $\lambda_1$ is non-zero (proven analytically in Sec.~\ref{sec:2.2}),
so the Wasserstein space is exactly one-dimensional.
For $p=2$, the squared distance $W_2^2$ equals the squared $L^2$ norm between quantile
functions as we can see from Eq.~\eqref{WpCDF}, so the Gram matrix is guaranteed to be positive-semidefinite and
all $N-1$ non-trivial eigenvalues are positive, giving $D_{\min}=N-1=15$ unless some eigenvalues happen to vanish.
The full-rank result for $p=2$ reflects the fact that each Husimi Q quantile
function $Q_n$ carries distinct shape information not captured by $Q_m$ for $m\neq n$;
this is a specific property of the Husimi Q-distribution of the harmonic oscillator.

\item[(ii)] \textbf{$\lambda_3<0$ for $p\geq 3$.}
For $p>2$, optimal transport distances are no longer equivalent to $L^2$ norms,
and Euclidean embedding is generically impossible.
For the Husimi Q-distribution we find numerically that $\lambda_3$ turns
negative at $p=3$ and its magnitude grows monotonically with $p$
(Table~\ref{tab:lam3}), confirming that no Euclidean embedding exists for $p\geq 3$.
This property reflects the fact that the embedding fails even at the minimal number of states $N=4$, which can be confirmed numerically.\footnote{Any three-point metric space can always be isometrically embedded in $\mathbb{R}^2$ (since any triangle is realizable in the Euclidean plane), so the minimal non-embeddable subset must contain at least four points. This also guarantees $\lambda_1, \lambda_2 > 0$ for the four-point Gram matrix.}

\begin{table}[h]
    \centering
    \begin{tabular}{c|cc}
     $p$ & $\lambda_3$ & $|\lambda_3|/\lambda_1$ \\
    \hline
    $3$ & $-8.5\times10^{-6}$ & $6.5\times10^{-7}$ \\
    $4$ & $-2.8\times10^{-5}$ & $2.1\times10^{-6}$ \\
    $5$ & $-5.1\times10^{-5}$ & $3.9\times10^{-6}$ \\
    $6$ & $-7.4\times10^{-5}$ & $5.7\times10^{-6}$ \\
    $7$ & $-9.6\times10^{-5}$ & $7.4\times10^{-6}$ \\
    $8$ & $-1.18\times10^{-4}$ & $9.0\times10^{-6}$ \\
    \end{tabular}
    \caption{The third eigenvalue $\lambda_3$ of the Gram matrix $B$ for the Husimi
    Q-distribution with $N=16$ states and $p=3,\dots,8$.
    %, computed at \texttt{WorkingPrecision}~$=60$.
    $\lambda_3$ is negative for all $p\geq 3$, and its  magnitude relative to $\lambda_1$
    grows monotonically with $p$.}
    \label{tab:lam3}
\end{table}

\item[(iii)] \textbf{$D_{\mathrm{eff}}\approx 1$ for $p=2$.}
Although $D_{\min}=N-1$ for $p=2$, the effective dimension
$D_{\mathrm{eff}}=1.000209$ is extremely close to unity,
reflecting a strong hierarchy $\lambda_1\gg\lambda_2\gg\cdots$ among the eigenvalues.
The dominant eigenvalue $\lambda_1$ captures the \emph{shift} of the Husimi Q-distribution
along the energy axis as $n$ increases---the same one-dimensional structure
that makes $p=1$ exactly rank-1.
The remaining eigenvalues encode the change in \emph{shape} beyond a pure shift,
and they are exponentially suppressed.
Figure~\ref{fig:spectrum} shows the eigenvalue spectrum for $N=16$, $p=2$:
the data approximately fall on a straight line in the log plot,\footnote{%
The coefficient $5.46$ in~\eqref{eq:spectrum_fit} is the extrapolated intercept of the log-linear fit; the dominant eigenvalue $\lambda_1=12.9936$ lies slightly above this trend.
The dominant eigenvalue $\lambda_1$ (the $k=0$ term), which controls the overall scale of the spectrum, can be evaluated analytically in terms of the Wasserstein coordinate $Z(n)$ defined in \eqref{Zn}:
$\lambda_1 \approx \sum_{n=0}^{N-1}(Z(n) - \bar{Z})^2$,
where $\bar{Z} = N^{-1}\sum_{n=0}^{N-1} Z(n)$.
For $N=16$ this gives $12.9893$, in agreement with the numerical value $\lambda_1 = 12.9936$ to within $0.03\%$. It can be also shown that $\lambda_1$ is estimated as $\sum_{n=0}^{N-1}(Z(n) - \bar{Z})^2 \sim N^2/18$ for $N\to\infty$.}
\begin{equation}
    %\log_{10}\lambda_k \approx 0.737 - 2.643\,k
    \lambda_k \approx 5.46 \times 10^{-2.64\, (k-1)}
    \qquad (k=1,\dots,N-1),
    \label{eq:spectrum_fit}
\end{equation}
with the coefficient of determination $R^2=0.997$,
corresponding to a ratio $\lambda_k/\lambda_{k+1}\approx 440$ per step.\footnote{The exponent $2.64$ in~\eqref{eq:spectrum_fit} is specific to $N=16$ and decreases monotonically with $N$.
%High-precision computations (WorkingPrecision 40--60) give
For different values of $N$, the exponent is numerically estimated as
$%\alpha_{10}(N)\approx 
2.78,\,2.71,\,2.65,\,2.64$
for $N=6,\,9,\,13,\,16$, respectively.
%Whether $\alpha_{10}(N)$ converges to a finite limit as $N\to\infty$ cannot be established from $N\le 16$ alone.
Further analysis may be necessary to determine whether the exponent converges to a finite value as $N$ increases.}
This exponential suppression is the reason why the $W_2$ and $W_1$ distance matrices
are numerically almost identical in Table~\ref{tab:ev2}: both distances are dominated
by the same one-dimensional shift mode.

The key distinction between $p=1$ and $p=2$ is therefore qualitative:
for $p=1$, the one-dimensional geometry is \emph{exact}
(the non-leading eigenvalues vanish identically),
while for $p=2$ it holds only \emph{approximately}
(they are non-zero but exponentially small).
This difference is reflected in $D_{\min}=1$ versus $D_{\min}=N-1$,
even though $D_{\mathrm{eff}}\approx 1$ in both cases.
The choice $p=1$ is therefore preferred if one demands an exact
one-dimensional holographic geometry.

\begin{figure}[h]
  \centering
  \includegraphics[width=0.56\linewidth]{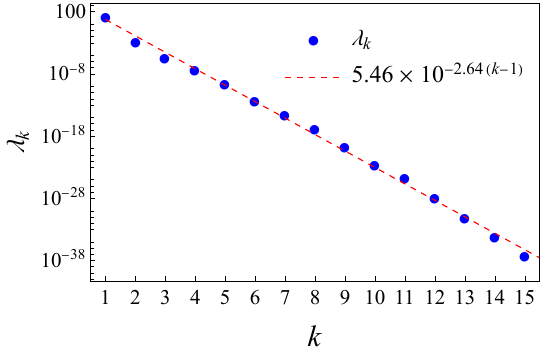}
  \caption{Eigenvalue spectrum of the MDS Gram matrix $B$ for $p=2$, $N=16$ states, where $\lambda_{16}=0$ is omitted.
%  (computed at \texttt{WorkingPrecision}~$=60$; $k=0,\dots,14$ in descending order;
%  the structural zero $\lambda_{15}=0$ is excluded).
  The dashed line is the fit $\lambda_k = 5.46\times 10^{ - 2.64\,(k-1)}$.}
  \label{fig:spectrum}
\end{figure}

\end{enumerate}
% [v2-CHANGE-END]

%\bibliographystyle{plain}
\bibliographystyle{JHEP}
\bibliography{For_paper1}

%\begin{thebibliography}{99}
%\end{thebibliography}

\end{document}